\documentstyle[12pt]{article}

\def\degres{\mbox{$^\circ$}}      

\def\gsim{\mbox{$\stackrel{_>}{_\sim}$}} 
\def\lsim{\mbox{$\stackrel{_<}{_\sim}$}} 

\def\bec{\begin{center}}
\def\eec{\end{center}}

\def\beq{\begin{equation}}
\def\eeq{\end{equation}}

\renewcommand{\frac}[2]{{{\displaystyle #1}\over{\displaystyle #2}}}

\def\MeV{\mbox{ MeV}}

\def\BeSv{\mbox{$^{7}$Be}}     
\def\BHt{\mbox{$^{8}$B}}

\def\calA{\mbox{$\cal A$}}     

\def\calR{\mbox{$\cal R$}}
\def\calS{\mbox{$\cal S$}}

\def\hath{\mbox{${\hat{h}}$}}

\def\dms{\mbox{$\Delta m^2$}}     

\def\SdTvS{\mbox{$\sin^2 2\theta_V$}}
\def\STvS{\mbox{$\sin^2 \theta_V$}}

\def\GF{\mbox{$G_F$}}               

\def\numt{\mbox{$\nu_{\mu(\tau)}$}} 
\def\nue{\mbox{$\nu_e$}}            
\def\nus{\mbox{$\nu_s$}}            
\def\rhoR{\mbox{$\rho_r$}}      

\def\Ye{\mbox{$Y_e$}}                 
\def\YeCore{\mbox{$Y_e(core)$}}       

\def\EDms{\mbox{$E_\nu/\Delta m^2$}} 
\def\Enu{\mbox{$E_\nu$}}         
\def\Te{\mbox{$T_e$}}            
\def\TeTh{\mbox{$T_{e,th}$}}     
\def\SeZr{\mbox{$\calS_{0}$}}       
\def\Ses{\mbox{$\calS^s$}}          
\def\SeD{\mbox{$\calS^D$}}          
\def\Res{\mbox{$\calR^s$}}          
\def\ReZr{\mbox{$\calR_{0}$}}       
\def\ReD{\mbox{$\calR^D$}}          
\def\AsymPs{\mbox{$\calA_P^s$}}     
\def\AsymPN{\mbox{$\calA_P^N$}}     
\def\AsymPC{\mbox{$\calA_P^C$}}     
\def\AsymSs{\mbox{$\calA_{D-N}^s$}} 
\def\AsymSN{\mbox{$\calA_{D-N}^N$}} 
\def\AsymSC{\mbox{$\calA_{D-N}^C$}} 
\def\AsymSM{\mbox{$\calA_{D-N}^M$}} 
\def\AsymRs{\mbox{$A_{D-N}^s$}}     
\def\AsymRN{\mbox{$A_{D-N}^N$}}     
\def\AsymRC{\mbox{$A_{D-N}^C$}}     
\def\AsymRM{\mbox{$A_{D-N}^M$}}     
\def\AsymRNCM{\mbox{$A_{D-N}^{N,C,M}$}}   
\def\Ps{\mbox{${\bar{P}}_\odot$}}       
\def\PTot{\mbox{$P_{\oplus}$}}          
\def\PTots{\mbox{$P^s_\oplus$}}         
\def\PeTw{\mbox{$P_{e2}$}}              
\def\APeTw{\mbox{$<\PeTw>$}}            
\def\APeTws{\mbox{$<\PeTw>^s$}}         
\def\APeTwN{\mbox{$<\PeTw>^{N}$}}       
\def\APeTwC{\mbox{$<\PeTw>^{C}$}}       
\def\APeTwM{\mbox{$<\PeTw>^{M}$}}       
\def\DAY{\mbox{\em{Day}}}                        
\def\night{\mbox{\em{Night}}}                    
\def\core{\mbox{\em{Core}}}                      
\def\mantle{\mbox{\em{Mantle}}}                  
\def\dseedEe{\mbox{$\frac{d\, \sigma_{\nu_e  } (\Te,E_\nu)}{d\,\Te}$}}
\def\daynight{D-N}                        
\def\SK{Super - Kamiokande}

\def\deg{\degres}  

\def\CdTv{\mbox{$\cos 2 \theta_V$}}

\def\ltap{\ \raisebox{-.4ex}{\rlap{$\sim$}} \raisebox{.4ex}{$<$}\ }
\def\gtap{\ \raisebox{-.4ex}{\rlap{$\sim$}} \raisebox{.4ex}{$>$}\ }

\def\ltap{\lsim}  

\def\electron{\mbox{$e^-$}}

\def\TabEventRates{I}  
\def\TabAsymmetries{S II}
\def\TabAsymmetriesMeV{S III}

\begin{document}
\sloppy


\vspace{0.5cm}
%
%
{\normalsize
\begin{flushright}
\begin{tabular}{l}
SISSA --154/97/EP\\
arch-iv/9803244
\end{tabular}
\end{flushright}
}
\vspace{1cm}

\begin{center}
{\Large
A Study of the Day - Night Effect for the \SK\  Detector: III.
The Case of Transitions into Sterile Neutrino.
}
\end{center}

\begin{center}
M. Maris $^{\mbox{a,b)}}$ and 
S. T. Petcov 
\footnote{
Also at: 
Institute of Nuclear
Research and Nuclear Energy, Bulgarian Academy of Sciences,
1784 Sofia, Bulgaria.} $^{\mbox{b,c)}}$ \\
$^{\mbox{a)}}$ OAT: Osservatorio Astronomico di Trieste, Trieste, Italy.\\
$^{\mbox{b)}}$ Scuola Internazionale Superiore di Studi Avanzati,
Trieste, Italy.\\
$^{\mbox{c)}}$ INFN - Sezione di Trieste, Trieste, Italy.\\
\end{center}

\bec
\abstract{
\setlength{\baselineskip}{16pt}
\noindent  Using the results of a high precision 
calculation of the solar neutrino survival probability
for Earth crossing neutrinos in the case of 
MSW $\nu_e \rightarrow \nu_s$\ transition
solution of the solar neutrino problem, we derive 
predictions for the one-year averaged
day-night (D-N) asymmetries in the deformed  
recoil-e$^-$ spectrum and in the energy-integrated
event rate due to the solar neutrinos, to be measured with the
Super - Kamiokande detector.
The asymmetries are calculated for three event samples, 
produced by solar $\nu_e$\ crossing 
the Earth mantle only, the core,
and the mantle only + the core (the full night sample),  
for a large set of representative values 
of the MSW transition parameters
$\Delta m^2$ and $\sin^2 2\theta_V$
from the ``conservative'' 
$\nu_e \rightarrow \nu_s$ solution regions,
obtained by taking into account the possible
uncertainties in the predictions for the
$^{8}$B and $^{7}$Be neutrino fluxes.
The effects of the uncertainties in the value
of the bulk matter density and in the chemical composition 
of the Earth core on the predictions for the 
D-N asymmetries are investigated. 
The dependence of the D - N effect related observables 
on the threshold recoil - e$^-$\ kinetic energy, $T_{e,th}$, 
is studied. It is shown, in particular, that for 
$\sin^2 2\theta_V \leq 0.030$\ the one
year average D-N - asymmetry in the sample of events due to 
the core-crossing neutrinos is larger than the asymmetry
in the full night  sample 
by a factor which, depending on the solution value of
\dms, can be $\sim (3 - 4)$ ($\dms ~< 5\times 10^{-6}~{\rm eV^2}$)
or $\sim (1.5 - 2.5)$ 
($5\times 10^{-6}~{\rm eV^2}~ 
\ltap ~\dms ~\ltap 8\times 10^{-6}~{\rm eV^2}$). 
We find, however, that at small mixing angles
$\sin^2 2 \theta_V ~\ltap~ 0.014$, 
the D-N asymmetry in
the case of solar $\nu_e \rightarrow \nu_s$\ transitions is 
considerably smaller than if the transitions were into an active
neutrino, $\nu_e \rightarrow \nu_{\mu(\tau)}$.
In particular, a precision better than 1\% in the measurement 
of any of the three one year averaged D-N asymmetries considered
by us would be required to test the small 
mixing angle nonadiabatic $\nu_e \rightarrow \nu_s$ solution
at $\sin^2 2\theta_V ~\ltap ~0.01$.
For $0.0075~ \ltap~ \sin^2 2\theta_V ~\leq~ 0.03$
the magnitude of the D-N asymmetry in the sample
of events due to the core-crossing neutrinos is
very sensitive 
to the value of the electron number fraction
in the Earth core, $Y_e(core)$: a change of 
$Y_e(core)$ from the standard value of 0.467 to 
the conservative upper limit of 0.50 can lead to an increase
of the indicated asymmetry by a factor
of $\sim (3 - 4)$. Iso - (D-N) asymmetry contours in the 
$\Delta m^2 - \sin^2 2 \theta_V$\ plane for the
Super-Kamiokande  detector are derived in 
the region $\sin^2 2\theta_V \geq 10^{-4}$
for the three event samples studied
for  $T_{e,th} = 5~{\rm MeV ~and~7.5~MeV}$, and
in the case of the samples due to the
 core crossing and (only mantle crossing + 
core crossing) neutrinos - for $Y_e(core) = 0.467~{\rm and~} 0.50$. 
The possibility to discriminate between the 
$\nu_e \rightarrow \nu_s$ and 
$\nu_e \rightarrow \nu_{\mu(\tau)}$ solutions
of the solar neutrino problem by performing high precision
D-N asymmetry measurements is also discussed.
 }
\eec

\newpage
\section{Introduction}

\indent In the present article we continue the systematic study of the 
\daynight\ effect for
the \SK\ detector began in refs. \cite{ArticleI}\ and \cite{ArticleII}. 
Assuming that the solar neutrinos undergo two - neutrino MSW 
$\nue \rightarrow \numt$\ 
transitions in the Sun, and that these transitions are at the origin of the 
solar neutrino deficit, we have performed in 
\cite{ArticleII}\ a high - 
precision calculation of the one - year averaged solar \nue\ survival 
probability for Earth crossing neutrinos, 
$\PTot(\nue\rightarrow\nue)$, reaching the \SK\ detector.
The probability $\PTot(\nue\rightarrow\nue)$\ 
was calculated by using, in particular, 
the elliptical orbit approximation (EOA) 
to describe the movement of the Earth around the Sun. 
Results for $\PTot(\nue\rightarrow\nue)$\ as a function 
of $\Enu/\dms$, $\Enu$\ and \dms\ 
being the neutrino energy and the neutrino mass 
squared difference, have been 
obtained for neutrinos crossing the Earth mantle only, 
the core, the inner 2/3 of 
the core and the mantle + core (full night) for a 
large representative set of values of 
\SdTvS, where $\theta_V$\ is the neutrino mixing angle in vacuum, from the 
''conservative'' MSW solution region in the 
\dms\ - \SdTvS\ plane, derived by taking 
into account the possible uncertainties in the fluxes 
of \BHt\ and \BeSv\ neutrinos (see, e.g., ref. \cite{SPnu96,KPUNPUB96}; 
for earlier studies see ref. \cite{KS94}).

 We have found in \cite{ArticleII}, in particular, that for 
$\SdTvS \leq 0.013$\ the one - year averaged \daynight\ asymmetry
\footnote{A rather complete list of references on the \daynight\
effect is given in ref. \cite{ArticleI}; for relatively recent discussions 
of the effect see, e.g., 
 \cite{
   Hata:Langacker:1994Earth,
   Baltz:Weneser:1994,
   Gelb:Kwong:Rosen:1996,
   Lisi:Montanino:1997,
   Bahcall:Krastev:1997,
   Hata:1997}
}
in the probability $\PTot(\nue\rightarrow\nue)$\ 
for neutrinos crossing the 
Earth core can be larger than the asymmetry in the probability for 
(only mantle crossing + core crossing) 
neutrinos by a factor of up to six.
The enhancement is even larger for neutrinos crossing the inner 2/3 of 
the core. We have also pointed out to certain subtleties in the calculation 
of the time averaged \nue\ survival probability 
$\PTot(\nue\rightarrow\nue)$\
for neutrinos crossing the Earth, which 
become especially important when $\PTot(\nue\rightarrow\nue)$\ is 
computed, for instance, for the core crossing neutrinos only~
\footnote{For further details concerning the technical 
aspects of the calculations see 
ref. \cite{ArticleII} as well as ref. \cite{NU4DN}.}. 

 The results obtained in \cite{ArticleII}\ were used in \cite{ArticleI}\ to
investigate the \daynight\ asymmetries in the spectrum of the recoil 
electrons from the reaction $\nu + e^- \rightarrow \nu + e^-$\ caused 
by the \BHt\ neutrinos and in the energy-integrated event rate, 
to be measured by the \SK\ experiment. 
We have computed in \cite{SKDNII:spectrum}\ the \daynight\ asymmetry in 
the recoil-$e^-$\ spectrum for the 
same large set of representative values of \dms\ and \SdTvS\ from the 
``conservative'' MSW solution region for which results in 
\cite{ArticleII}\ have been presented. The \daynight\ asymmetry in the 
$e^-$-spectrum was found for neutrinos crossing the Earth mantle only, 
the core  and the mantle + core.
In \cite{ArticleI} we have included only 12 representative plots 
showing the 
magnitude of the \daynight\ asymmetry in the recoil-e$^{-}$ spectrum 
to be expected in the case of the 
two - neutrino MSW $\nue \rightarrow \numt$\ 
transition solution of the solar neutrino problem.
The spectrum asymmetry for the sample of events due to core crossing 
neutrinos only was found to be strongly enhanced for 
$\SdTvS~\lsim~0.013$\ with respect to the analogous asymmetries for the 
mantle and for the (only mantle crossing + core crossing) neutrinos. 
We presented in \cite{ArticleI} also detailed results for 
the one-year averaged \daynight\ 
asymmetry in the \SK\ signal for the indicated three samples of events.
We have found that indeed for  
$\sin^22\theta_{V}\leq 0.013$\ the asymmetry in the sample 
corresponding to core crossing neutrinos can be larger than the asymmetry 
in the sample produced by only mantle crossing or by 
(only mantle crossing + core crossing ) neutrinos by a factor of up to six.
We have investigate in \cite{ArticleI} 
the dependence of the D-N asymmetries in the 
three event samples defined above on the threshold
e$^{-}$ kinetic energy being used for the event selection.
The effect of the uncertainties in the Earth matter density
and electron number density distributions on the predicted values
of the D-N asymmetries were studied as well. 
We derived also iso - (\daynight) asymmetry contours in the region of 
$\SdTvS~\gsim~10^{-4}$\ in the \dms - \SdTvS\ plane for the signals in 
the \SK\ detector, produced by neutrinos crossing the mantle only, the core 
and the mantle + core (full night sample). 
The iso-asymmetry contours for the sample of events due to
core-crossing neutrinos were obtained for two values of the
fraction of electrons, $Y_e$, in the core: for $Y_e (core) = 0.467$ and 0.500
\footnote{Results for other D-N effect related observables,
as the D-N asymmetry in the zenith angle distribution of the events
and in the mean recoil-e$^{-}$ energy, have been obtained 
in refs. \cite{Lisi:Montanino:1997, Bahcall:Krastev:1997}, where
iso - (D-N) asymmetry contour plots for the 
full night (i.e., mantle + 
core) signal for the \SK\ detector for one value of $Y_e(core) = 0.0467$
have been presented as well.}.
The results derived in \cite{ArticleI} 
confirmed the conclusion drawn in \cite{ArticleII}\ 
that the \SK\ detector will be able to probe   
not only the large mixing angle adiabatic 
solution region, but also 
an important and substantial part of 
the $\SdTvS~\lsim~0.014$\ nonadiabatic
region of the MSW $\nue \rightarrow \numt$\ transition
solution of the solar neutrino problem. We have found, in particular,
that in a large sub-region of the ``conservative'' 
nonadiabatic solution region
located at $\SdTvS~\lsim~0.0045$, the D-N asymmetry in the 
sample of events due to the core-crossing neutrinos only is negative
and has a value in the interval (-1\%) - (-3\%).

 In the present article we realize the same program of studies for the
Super-Kamiokande detector
for the alternative possibility of solar neutrinos undergoing 
two-neutrino matter-enhanced transitions in the Sun and in 
the Earth into a sterile 
neutrino, \nus. As is well-known, the solar $\nu_e$ 
matter-enhanced transitions into a sterile 
neutrino, $\nu_e \rightarrow \nu_s$, provide one of the possible
neutrino physics solutions of the solar 
neutrino problem (see, e.g., 
\cite{KPNPB95,Hata:Langacker:1994Earth,KPL96,SPnu96}). 
The reference solution region in the $\Delta m^2 - \sin^22\theta$ plane,
i.e., the region obtained (at 95\% C.L.) 
by using the predictions of the reference 
solar model of Bahcall and Pinsonneault from 1995 with 
heavy element diffusion \cite{BP95} ~(BP95) for the different 
components of the solar neutrino flux 
(pp, pep, $^{7}$Be, $^{8}$B and CNO), lies within the bounds:
\vspace{-1.2cm}
\bec\beq
2.8\times 10^{-6}~{\rm eV}^2 ~\ltap ~\Delta m^2
     ~\ltap ~7.0\times 10^{-6}~{\rm eV}^2,~~
\eeq\eec
\vspace{-1.0cm}
\bec\beq
4.8\times 10^{-3}~ \ltap ~\sin^22\theta ~\ltap ~1.4\times 10^{-2}~.
\eeq\eec
\noindent The reference solution is of the small mixing angle
nonadiabatic type. A reference large mixing angle (adiabatic)
solution (present in the case of 
$\nu_{e} \rightarrow \nu_{\mu (\tau)}$ transitions)
is practically ruled out by the solar neutrino data
\cite{KPNPB95,KPL96,SPnu96}. If we allow for
possible uncertainties in the predictions for the fluxes
of the $^{8}$B and $^{7}$Be neutrinos \cite{KPL96}, the solar neutrino data 
is described in terms of the hypothesis of 
$\nu_e \rightarrow \nu_s$ transitions
for larger ranges of values of the parameters 
$\Delta m^2$ and $\sin^22\theta$, belonging to the intervals:
\vspace{-1.2cm}
\bec\beq
2.8\times 10^{-6}~{\rm eV}^2 ~\ltap ~\Delta m^2
     ~\ltap ~8.0\times 10^{-6}~{\rm eV}^2,~~
\eeq\eec
\vspace{-1.0cm}
\bec\beq
8.0\times 10^{-4} ~\ltap ~\sin^22\theta ~\ltap ~3.0\times 10^{-2}~.
\eeq\eec
\vspace{-0.4cm}
\noindent and 
\vspace{-1.2cm}
\bec\beq
5.3\times 10^{-6}~{\rm eV}^2 ~\ltap ~\Delta m^2
     ~\ltap ~1.2\times 10^{-5}~{\rm eV}^2,~~
\eeq\eec
\vspace{-1.0cm}
\bec\beq
0.13 ~\ltap ~\sin^22\theta ~\ltap ~0.55~.
\eeq\eec
\noindent The ``conservative'' solution regions lying within the
bounds determined by eqs. (3) - (4), and eqs. (5) - (6) have been obtained 
by treating the $^{8}$B neutrino flux as a free 
parameter in the relevant analysis of the solar neutrino data, while 
the the $^{7}$Be neutrino flux was assumed to have a value in the interval
$\Phi_{Be} = (0.7 - 1.3)~\Phi^{BP}_{Be}$, where 
$\Phi^{BP}_{Be}$ is the flux in the reference solar model \cite{BP95}.
For values  of $\Delta m^2$ and $\sin^22\theta$ from the
solution regions (5) and (6) the 
$\nu_e \rightarrow \nu_s$ transitions of $^{8}$B neutrinos 
having energy $E_{\nu} \geq 5~{\rm MeV}$ are adiabatic.
However, this adiabatic solution is possible 
only for large values of the 
$^{8}$B neutrino flux \cite{KPL96}, 
$\Phi_{B} \cong (2.5 - 5.0)~\Phi^{BP}_{B}$,
$\Phi^{BP}_{B}$ being the reference model flux \cite{BP95}. 
Such values of $\Phi_{B}$ 
seem totally unrealistic from the point of view of the
contemporary solar models and we consider the indicated 
adiabatic solution here for completeness.
It should be added that in deriving the conservative 
solution regions represented by eqs. (3) - (4) and eqs. (5) - (6)
the limit on the D-N effect derived in 
\cite{Hata:Langacker:1994Earth} 
on the basis of the data obtained in the Kamiokande II and III experiments
\cite{KamDN} was utilized. 

   Let us note that the 
preliminary result on the D-N effect
from the Super-Kamiokande experiment after approximately one year
(374.2 days) of data taking reads \cite{SKSB97}:
\vspace{-0.6cm}
\bec\beq
     \bar{A}^{SK}_{D-N} \equiv \, \frac{\bar{R}^{D} - \bar{R}^{N}}
{\bar{R}^{D} + \bar{R}^{N}} =  - 0.031 \pm 0.024 \pm 0.014, 
\eeq\eec
\vspace{0.2cm}
\noindent where $\bar{A}^{SK}_{D-N}$ is the average 
energy integrated D-N asymmetry and 
$\bar{R}^{D}$ and $\bar{R}^{N}$ are the observed average 
event rates caused by the solar neutrinos during the day and during
the night in the Super-Kamiokande detector in the
period of data taking. The first error in eq. (7) is 
statistical and the second error is systematic.
The data were obtained with a recoil$-e^{-}$ threshold energy 
$E_{e,th} = 6.5~{\rm MeV}$. 

   In addition of performing i) detailed high precision calculations
of the D-N asymmetries in the recoil-e$^{-}$ spectrum and ii) of the
energy-integrated D-N asymmetries for the three samples of events
(due to core crossing, only mantle crossing and only mantle + core crossing
neutrinos), and of studying iii) the effects of the
recoil-e$^{-}$ energy threshold variation and iv) of the uncertainties
in the chemical composition and matter density of the Earth's core
on the calculated D-N effect related observables, 
we also analyze qualitatively the possibility
to distinguish between the two
solutions of the solar neutrino problem involving
matter-enhance transitions of the solar $\nu_e$ respectively 
into active neutrinos
and into sterile neutrinos,  
$\nu_e \rightarrow \nu_{\mu (\tau)}$ and
$\nu_e \rightarrow \nu_s$,
by performing high-precision D-N asymmetry measurements. 

\vspace{-0.3cm}  
\section{Specific Features of the Earth Matter 
Effect for Solar Neutrino Transitions Into Sterile Neutrino}

\indent  In the present study we utilize the same high precision methods of
calculation of i) the position of the Sun (at a given time of the year t)
with respect to the Super-Kamiokande detector (elliptical orbit
approximation which accounts for the change with t of the Earth 
orbital velocity as well), ii) the solar $\nu_e$ survival probability in
the Sun, $\bar {P}_{\odot} (\nu_e \rightarrow \nu_e)$, 
and iii) the $\nu_2 \rightarrow \nu_e$ transition probability
in the Earth (see, e.g., \cite{ArticleII}), $P_{e2}$,
$\nu_2$ being the heavier
of the two vacuum mass-eigenstate neutrinos, which were used in our 
studies of the D-N effect for the Super-Kamiokande detector 
in the case of solar $\nu_e$ transitions into active neutrinos, 
$\nu_{\mu (\tau)}$. They are described in detail in refs. 
\cite{ArticleI,ArticleII}. In the present Section we discuss the
differences between the $\nu_2 \rightarrow \nu_e$ transitions in the 
Earth in the cases of the  $\nu_e \rightarrow \nu_s$ and 
$\nu_e \rightarrow \nu_{\mu (\tau)}$  
transition solutions of the solar neutrino problem, which are relevant for
our further analysis and also allow to understand qualitatively the
differences between the results on the D-N effect in the two cases.

  As we have indicated above, the calculation of the Earth 
effect related observables 
in the case of interest requires the knowledge of the probabilities
$\bar {P}_{\odot} (\nu_e \rightarrow \nu_e)$ and $P_{e2}$ which account
for the matter-enhanced transitions of the solar 
neutrinos in the Sun and in the Earth, respectively. 
The neutrino effective potential in matter which enters in
the system of evolution equations describing the 
$\nu_e \rightarrow \nu_s$ transitions is \cite{Langacker:1986}:
\vspace{-0.8cm}
\bec\beq\label{eq:sterile:potential}
 V_{s}(x) = \sqrt{2} \GF (n_e(x) - \frac{1}{2} n_n(x)),   
\eeq\eec

\noindent where \GF\ is the Fermi constant 
and $n_e(x)$ and $n_n(x)$\ are the electron 
and neutron number densities at a given point $x$\ of the neutrino
trajectory. 

   For the Sun we use 
the electron and neutron number density distributions given by the 
standard solar model \cite{BP95} with heavy 
element diffusion \footnote{All solar models compatible with the currently
existing observational constraints (helioseismological and other) 
give practically the same electron and neutron number density 
distributions in the Sun.}.
The ratio $0.5~n_n/n_e$ changes from approximately 0.22 
in the central part to 0.08 in most of the Sun outside the 
neutrino (energy) production region (see, e.g., \cite{BP95}), 
so the presence of the
neutron number density term in $V_s(x)$ does not play important role
in the solar neutrino transitions in the Sun, although 
it has to be taken into account in the calculation of 
$\bar {P}_{\odot} (\nu_e \rightarrow \nu_e)$.
 
 The electron and neutron number densities in the Earth are 
directly related to 
the local matter density, $\rho(r)$, where $r$ is the distance from the Earth 
center, and the local chemical composition which is accounted for 
by the electron fraction number, $Y_e(r)$: $Y_e(r) = n_e(r)/(n_n(r) + n_p(r))$,
$n_p(r)$ being the proton number density, $n_p(r) = n_e(r)$.
As in \cite{ArticleI, ArticleII}, 
the Stacey model from 1977 \cite{Stacey:1977} is utilized as a 
reference Earth model 
in the calculations performed for the present study.
The Earth radius in the Stacey model is $R_{\oplus} = 6371~$km.
As in all Earth models known to us, the density distribution
is spherically symmetric and there are two major density structures - 
the core and the mantle, and 
a large number of substructures (shells or layers). The core 
has a radius $R_c = 3485.7~$km,
so the Earth mantle depth is approximately $R_{man} = 2885.3~$km.  
The mean matter densities in the core and in the mantle read, respectively:
$\bar{\rho}_c \cong 11.5~{\rm g/cm^3}$ and 
$\bar{\rho}_{man} \cong 4.5~{\rm g/cm^3}$.
Let us note that the density distribution in the 1977 Stacey model
practically coincides with the density distribution
in the more recent PREM model \cite{PREM81}.

   The onion like (shell) structure of the Earth is 
reproduced in the Stacey model 
by a set of polynomial functions describing the 
radial change of the matter density $\rho(r)$.
Assuming the chemical composition in each geological structure (core, mantle, 
etc.) to be to a good approximation constant, the effective number 
density of scattering centers in
eq. (1), $(n_e - 1/2n_n)$, 
is obtained for each shell by simply rescaling the matter 
density of the $i-th$ shell, $\rho_i(r)/m_N$,  $m_N$ being the nucleon mass,
by the factor:
\vspace{-0.8cm}
\bec\beq
\frac{1}{2} (3 Y_{e,i} - 1), \,\,\, i = \mbox{geological shell 
index}
\eeq\eec
\noindent
where $Y_{e,i}$ is the electron fraction in the i-th shell.
Let us note that due to the factor $3/2$ in front of $Y_e$ in eq. (2),
the $\nu_e \rightarrow \nu_s$ transitions are more sensitive to the
uncertainty in the value of $Y_e$ in a given Earth shell than the
$\nu_e \rightarrow \nu_{\mu (\tau)}$ transitions.

  If we denote by $V_{a}(x)$ the neutrino effective potential in
matter relevant for the solar $\nu_e$ transitions into an active neutrino,
$V_{a}(x) = \sqrt{2} \GF n_e(x)$, we always have $V_{s}(x) < V_{a}(x)$
in the case of interest. Actually, in the Earth 
\vspace{-1.0cm}
\bec\beq
  V_{s}(r) \cong \frac{1}{2}~ V_{a}(r)
\eeq\eec
\noindent due to the approximate isotopic symmetry 
($n_p(r) \cong n_n(r)$) and neutrality ($n_p(r) = n_e(r)$) of the Earth
matter. This has several important implications.
For a fixed neutrino energy $E_{\nu}$ and given 
$\Delta m^2$ and $\cos2\theta$, 
the resonance density for the 
$\nu_e \rightarrow \nu_s$ transitions,
\vspace{-0.8cm}
\bec\beq
\rho^{res}_{s} = 
\frac{\Delta m^2 \cos2\theta}{2E\sqrt{2} G_F \frac{1}{2}(3Y_e - 1)},
\eeq\eec
\noindent is approximately by a factor of two larger 
than the resonance density for the $\nu_e \rightarrow \nu_{\mu (\tau)}$ 
transitions, 
$\rho^{res}_{a} = \Delta m^2 \cos2\theta /(2E\sqrt{2} G_F Y_e)$: 
\vspace{-1.0cm}
\bec\beq
   \rho^{res}_{s} \cong 2~\rho^{res}_{a}. 
\eeq\eec
\noindent  Correspondingly, 
for a given $\rho(r)$ and $\cos2\theta$,
the resonance condition in the $\nu_e \rightarrow \nu_s$ case will be 
fulfilled for a value of  $E_{\nu}/\Delta m^2$, which is two times 
bigger than the analogous resonance $E_{\nu}/\Delta m^2$ value
for the $\nu_e \rightarrow \nu_{\mu (\tau)}$ 
transitions, 
$(E_{\nu}/\Delta m^2)^{res}_{s} \cong 2~(E_{\nu}/\Delta m^2)^{res}_{a}$.
Thus, the oscillation length in matter at resonance in the Earth,
$L^{res}_{m} = 4\pi (E_{\nu}/\Delta m^2)^{res}/\sin 2\theta$, 
for solar neutrino transitions into sterile neutrino 
will exceed the resonance oscillation length in matter 
for transitions into an active neutrino by a factor of two:
$L^{res}_{m,s} \cong 2~L^{res}_{m,a}$. 
At small mixing angles the probability $P_{e2}$ typically 
should have a maximum
at $\rho^{res} \cong \bar{\rho}$ ($\sin^22\theta_m \cong 1$,
$\theta_m$ being the mixing angle in matter)
independently of the type of the transition,
$\nu_e \rightarrow \nu_s$ or
$\nu_e \rightarrow \nu_{\mu (\tau)}$, where
$\bar{\rho}$ is the effective mean density along the 
neutrino trajectory in a given Earth density structure. 
For the neutrino trajectories for which the matter
effect is significant one has 
$\bar{\rho}_{c} \cong (10 - 12)~ {\rm g/cm^3}$ for the core
and $\bar{\rho}_{man} \cong (3.5 - 5.0)~ {\rm g/cm^3}$ for the mantle.

   The small mixing angle (SMA) 
$\nu_e \rightarrow \nu_s$ and $\nu_e \rightarrow \nu_{\mu (\tau)}$
transition solutions of the solar neutrino problem
take place roughly for the same values of
$\Delta m^2$ and $\sin^22\theta$ \footnote{The
$\nu_e \rightarrow \nu_s$ transition 
SMA solution region in the $\Delta m^2 - \sin^22\theta$ plane is shifted
on average by a factor of 1.2 to smaller values of 
$\Delta m^2$ with respect to the analogous region
of the $\nu_e \rightarrow \nu_{\mu (\tau)}$ solution
\cite{KPL96}.}. 
It is easy to convince oneself 
that for the energies of $^{8}$B neutrinos of interest,
$E_{\nu} \cong (5 - 14)~$MeV, and values of $\Delta m^2$ from
the region of the SMA solutions, 
$\Delta m^2 \cong (3 - 10)\times 10^{-6}~$eV$^{2}$, we have
for a resonance taking place 
in the mantle ($\rho^{res} \cong \bar{\rho}_{man}$):
$(2\pi R/L^{res}_{m})^2 \ll 1$. As a consequence 
at resonance (where $\rho^{res} = \bar{\rho}$ and $\sin^22\theta_m = 1$), 
as can be shown, one approximately has: 
$(P_{e2} - \sin^2\theta) \sim 0.5 (2\pi R/L^{res}_{m})^2$. Thus,
for the only mantle crossing neutrinos,
at small mixing angles ($\sin^22\theta~\lsim~ 0.03$) 
and for a given $\Delta m^2$, 
the Earth matter effect term  $(P_{e2} - \sin^2\theta)$
in the solar $\nu_e$ survival
probability $P_{\oplus}(\nu_e \rightarrow \nu_e)$
is approximately by a factor of (3.5 - 4.0) smaller
in the case of $\nu_e \rightarrow \nu_s$ transitions 
than if the solar $\nu_e$
underwent $\nu_e \rightarrow \nu_{\mu (\tau)}$ transitions.
This conclusion agrees very well with our 
numerical results for the probability $P_{e2}$ for solar neutrinos 
crossing only the mantle when they traverse the Earth.
For the core crossing neutrinos the analysis is
more complicated and the above simple estimate for the ratio of the 
maximal values of $(P_{e2} - \sin^2\theta)$ corresponding to
the $\nu_e \rightarrow \nu_s$ and 
$\nu_e \rightarrow \nu_{\mu (\tau)}$ transitions
has to be modified:
actually, the corresponding 
factor is smaller than the one for the only mantle
crossing neutrinos, being approximately equal to $(2.5 - 3.0)$.
Thus, on the basis of the above results we can expect 
that at small mixing angles the 
Earth matter effect   
in the solar $\nu_e$ survival probability, 
$P_{\oplus}(\nu_e \rightarrow \nu_e)$,
will be substantially smaller 
(roughly by a factor of $\sim (2.5 - 4.0)$)
for the $\nu_e \rightarrow \nu_s$ solution
of the solar neutrino problem
than for the $\nu_e \rightarrow \nu_{\mu (\tau)}$ one. 

 Further important implications of equation (5) for the magnitude of the
D-N effect if the solar $\nu_e$ undergo transitions into a sterile neutrino
will be discussed in the next Section. Here we would like to add
one more remark only.
Following the detailed discussion of the uncertainties in the knowledge
of the chemical composition and the bulk matter density
of the Earth core and the method chosen to effectively 
account for these uncertainties in \cite{ArticleI},
we use two values of $Y_e$ in the core
in the present analysis: the reference Earth model value
$Y_e(core) = 0.467$ which has been used also in refs.
\cite{ArticleI, ArticleII,Lisi:Montanino:1997,Bahcall:Krastev:1997}, 
and the value $Y_e(core) = 0.500$, which represents a conservative upper limit 
for $Y_e(core)$ \cite{CORE}. In all calculations performed for 
the present study the value of $Y_e = 0.49$ in the 
mantle has been utilized.  
\vspace{-0.3cm}
\section{The Time-Averaged Probabilities}

\indent  The relevant probabilities for the problem of interest, in
addition to the probability of
solar \nue\ survival in the Sun, \Ps, and
the instantaneous $\nu_2 \rightarrow \nue$\ transition 
probability in the Earth,\PeTw, are
the time averaged of \PeTw\ over the full night or part of it, \APeTws, 
where the index $s$ indicates over which time period of the night 
(or equivalently, over which neutrino trajectories 
in the Earth) \PeTw\ is averaged, 
and the time averaged \nue\ survival probability when the solar neutrinos 
cross the Earth, $P^{s}_{\oplus}(\nu_e \rightarrow \nu_e)$.
As in \cite{ArticleI,ArticleII}, we will consider three types 
of time averaging which
correspond to three different selections of events in the 
Super-Kamiokande detector: those due to neutrinos 
which cross only the mantle, due to the core crossing neutrinos, and the 
events collected during the full night. 
We shall call these event samples respectively \mantle, \core\ and \night\ 
samples and will denote them by $M$, $C$ and $N$: $s=M, C, N$.

It proves useful to study the \daynight\ asymmetry in the probability of
solar \nue\ survival: 
\vspace{-0.6cm}
\bec\beq
  \AsymPs (E_{\nu}/\Delta m^2) \equiv  \AsymPs = 
   2~\frac{\PTots - \Ps}{\PTots + \Ps}, \,\,\, s=M, C, N.
\eeq\eec

\noindent
The values of the asymmetry \AsymPs\ give an idea about the magnitude of 
the \daynight\ effect to be expected in the corresponding sample of events.

The calculation of the probabilities 
\APeTws\ and \PTots\ 
is performed using the methods described in \cite{ArticleII}. 
For a rather large set of values of \SdTvS\
our results for \APeTws, \Ps, \PTots\ and the probability asymmetries \AsymPs,
$s = N, M, C,$\ are presented graphically in Figs. 3.1 - 3.12.

  In Figs. 3.1a - 3.12a we show \APeTws, $s = N, M, C,$\ as a function of the
density parameter 
\vspace{-0.8cm}
\bec\beq
   \rho_r = \frac{\dms \CdTv}{2 \Enu \sqrt{2} \GF 0.25}
\eeq\eec
\vspace{0.2cm}
\noindent for fixed values of \SdTvS.
This parameter would coincide with the resonance density if \Ye\ were equal to
$1/2$ both in the mantle and in the core; it is equivalent to 
$E_{\nu}/\Delta m^2$, but gives an idea about the densities 
at which one has an enhancement of the
Earth matter effect. In sub-figures 3.1b - 3.12b, 
3.1c - 3.12c and 3.1d - 3.12d
the probabilities \Ps, \APeTws, \PTots\ (upper frames) and the related 
asymmetries \AsymPs\ (lower frames) are shown as functions of \EDms.

 The enhancement of \APeTws\
due to the Earth matter effect is clearly seen in the figures.
The dependence of \APeTws\ on \EDms\ at small and at 
large mixing angles is quite different. 
The most remarkable feature, as in the case of transitions 
into an active neutrino \cite{ArticleI,ArticleII}, is the enhancement 
at small mixing angles, $\SdTvS ~\ltap~ 0.03$,
of \APeTw\ for the core-crossing neutrinos, \APeTwC,
with respect to \APeTw\ for the only mantle crossing neutrinos, \APeTwM, 
and with respect to \APeTwN\
(core enhancement).
As a quantitative measure of the enhancement one can consider the ratio
of \APeTwC\ and \APeTwN\ at their maxima: $\max(\APeTwC)/\max(\APeTwN)$.
In the case of $\nu_e \rightarrow \nu_s$ transitions
one typically has
$3.3 ~\lsim~\max(\APeTwC)/\max(\APeTwN)~\lsim~3.8$\ 
for $0.001 \leq \SdTvS \leq 0.03$, 
which is of the same order of magnitude 
as in the case of $\nue \rightarrow \numt$\ transitions.
In contrast, at large mixing angles, $\SdTvS \sim (0.1 - 0.6)$, we have: 
$\max(\APeTwC)/\max(\APeTwN)~ \lsim~ 1.5$.
Thus, as like for the $\nue \rightarrow \numt$\ transitions,
the core enhancement is significant only at small mixing angles.

  At large mixing angles, $\SdTvS \sim (0.1 - 0.6)$, we note the existence
of a substantial 
negative \daynight\ asymmetry in the \core\ sample
(Figs. 3.11 - 3.12).  
Physically, a negative \daynight\ asymmetry means that the Earth effect 
suppresses the solar \nue\ flux instead of enhancing it, so that the 
solar neutrino signal is smaller at night than during the day.
At small mixing angles a negative \daynight\ effect occurs 
when the survival probability in the Sun, \Ps, is larger than 1/2.
In the case of $\nue \rightarrow \nus$\ transitions 
this happens at large mixing angles 
because for certain values of $\rho_r$ 
(or equivalently of \EDms) 
we have $\APeTwC < \sin^2\theta_V~$
\footnote{The negative \daynight\ effect 
takes place at large mixing angles also in the 
$\nue \rightarrow \numt$\ case, but it is much smaller in magnitude and 
the integration over the recoil-electron energy makes it negligible
\cite{ArticleI, ArticleII}.}.

  Let us discuss next briefly the behavior of 
the probabilities \APeTws\ and the
corresponding asymmetries \AsymPs\ as \SdTvS\ changes from 0.001 to 0.6.
For values of \SdTvS\ from the interval (0.001 - 0.006) there are
two clear ``peaks'' in the D-N asymmetry in the probabilities, 
\AsymPs, but 
the only physically relevant ``peak'' is the ``negative'' ($\AsymPs < 0$) one:
the ``positive'' ($\AsymPs > 0$) peak located at 
smaller values of \EDms\ (see Figs. 3.1 - 3.7) 
is caused \cite{ArticleII} essentially by the small value of the 
adiabatic minimum of the 
probability \Ps, $min~\Ps = \sin^2\theta_V$.
As $\sin^22\theta$ increases the negative ``peak''
in  \AsymPC\ (\AsymPN)\   
increases in absolute value and reaches a maximum approximately at 
$\sin^22\theta \cong 0.007$; when $\sin^22\theta$ increases further
its significance begins to diminish, 
as physically relevant positive ``peaks'' 
preceding it along the \EDms\ axis appear and become 
prominent. For $\sin^22\theta \approx 0.007$, the negative
``peak''  corresponds to $\AsymPC \approx -2.8\%$\, while 
the dominant positive one is approximately equal to 1\%; 
at $\sin^22\theta \approx 0.009$ the two ``peaks''
in \AsymPC\ are practically equal in absolute value.

  Actually, the dominant physically relevant ``positive''
peak appears in \AsymPC\ because for  
$0.006 ~\ltap ~\sin^22\theta ~\ltap ~0.01$ 
the region of resonance enhancement of 
\APeTwC\ is divided into two parts by the 
$\Ps = 1/2$\ zero Earth effect ($\AsymPs = 0$) line 
located in the $E_{\nu}/\Delta m^2-$region where \Ps\ is non adiabatic. 
As a consequence, one has in 
the resonance region, i.e., in the region of 
the absolute maximum of \APeTwC:
$\AsymPC < 0$ if $E_{\nu}/\Delta m^2 > (E_{\nu}/\Delta m^2)_0$ and
$\AsymPC > 0$ for  
$E_{\nu}/\Delta m^2 < (E_{\nu}/\Delta m^2)_0$, where 
$(E_{\nu}/\Delta m^2)_0$ is the point at which
$\Ps = 1/2$ and $\AsymPC = 0$. The contributions from the regions
of positive and negative \AsymPC\ 
to the energy-integrated D-N \core\ asymmetry can partially
compensate each other. The degree of compensation depends
on the value of $\Delta m^2$. 
For $0.014 ~\ltap ~\sin^22\theta ~\ltap ~0.03$
the $\Ps = 1/2$\ (i.e., zero Earth effect) line is outside the
region of resonance enhancement of \APeTwC\ and 
$\AsymPC > 0$ in this region.  
The maximal value of $|\AsymPC |$ (i.e., the value at the relevant 
negative ``peak'') is nearly  
constant for $0.004 \ltap \sin^22\theta \ltap 0.007$
because of the interplay between the change of \APeTwC\  
and the change of the position of the $\Ps = 1/2$\ line with respect to
the position of the \APeTwC\ resonance maximum (Figs. 3.4 - 3.6).

  It is important to note that the value
$(E_{\nu}/\Delta m^2)_0$, at which 
$\Ps = 1/2$ and the $\nu_e$ transitions in the Sun are nonadiabatic,
is practically the same for the $\nu_e \rightarrow \nu_s$ and
the $\nu_e \rightarrow \nu_{\mu (\tau)}$ transitions \cite{KPL96}.
However, since the dominating 
absolute maximum of \APeTwC\
occurs in the two cases for values of $E_{\nu}/\Delta m^2$ 
which differ by approximately a factor of two,
$(E_{\nu}/\Delta m^2)^{res}_{s} \cong 2~(E_{\nu}/\Delta m^2)^{res}_{a}$,  
the zero Earth effect point $(E_{\nu}/\Delta m^2)_0$
in the case of solar $\nu_e$ transitions into an active
neutrino ``leaves'' the
region of the absolute maximum of \APeTwC\ 
as $sin^22\theta$ increases at a considerably smaller value of 
$sin^22\theta$ than in the case of 
$\nu_e \rightarrow \nu_s$ transitions, namely at 
$sin^22\theta \cong 0.006$ \cite{ArticleII}. 
Therefore, in contrast to the possibility described above for the 
$\nu_e \rightarrow \nu_s$ transitions,   
a partial or complete compensation 
between the negative and positive 
contributions to the energy-integrated D-N \core\ asymmetry
can occur for the  
$\nu_e \rightarrow \nu_{\mu (\tau)}$ transitions
only at $sin^22\theta ~\ltap ~0.005$.  
This together with the observations made in the previous Section
implies that for values of $\sin^22\theta$  
from the interval
$0.006 ~\ltap ~\sin^22\theta ~\ltap ~0.03$,
in which the reference and 
most of the ``conservative'' regions of the SMA nonadiabatic
$\nu_e \rightarrow \nu_{s}$ and
$\nu_e \rightarrow \nu_{\mu (\tau)}$ solutions lie,
the energy-integrated D-N \core\ asymmetry 
corresponding to the $\nu_e \rightarrow \nu_{s}$ transitions
should be much smaller than
the asymmetry due to $\nu_e \rightarrow \nu_{\mu (\tau)}$
transitions.
Similar considerations apply to the 
\mantle\ and \night\ asymmetries as well.

   As \SdTvS\ increases from $\SdTvS \approx 0.007$,
the asymmetry \AsymPs\  steadily
increases for all the samples until \SdTvS\ reaches approximately 
the value of $\sim (0.1 - 0.2)$. At larger values the dominant 
positive ``peak'' in \AsymPC\ 
is followed, as $E_{\nu}/\Delta m^2$ increases, by a prominent 
negative ``peak'' and several smaller maxima. 
The negative ``peak'' is due to the inequality 
$\APeTwC < \STvS$ taking place in a specific region of values of 
$E_{\nu}/\Delta m^2$.
For $\SdTvS \cong 0.4$\,  the probability \daynight\ 
asymmetries for the \night\ 
and \core\ samples are roughly equal, the ratio $\APeTwC / \APeTwN$
varying between 1.0 and 1.5,
except in the narrow region of the indicated negative ``peak'' in \APeTwC.
As \SdTvS\ increases further we have $\APeTwC / \APeTwN < 1$\  
in a relatively large interval of values of \EDms\ (Fig. 3.12).

    Changing \YeCore\ from 0.467 to 0.50, which accounts to large extent 
for the uncertainties in the knowledge of the 
core chemical composition and density
(see \cite{ArticleI,CORE}) shifts the position of the
absolute (resonance) maximum of \APeTwC\ (\APeTwN) to a smaller 
by a factor of 1.25 value of $E_{\nu}/\Delta m^2$. 
As a consequence, the zero Earth effect point
$(E_{\nu}/\Delta m^2)_0$ falls outside the region of
resonance enhancement of 
\APeTwC\ (as $sin^22\theta$ increases) at smaller value 
of $sin^22\theta$ than in the 
$Y_e = 0.467$ case. Accordingly, for 
$0.008 ~\ltap ~sin^22\theta ~\ltap ~0.03$, the D-N \core\ asymmetry 
in the probability \AsymPC\ for $Y_e = 0.50$ is bigger
than the asymmetry for $Y_e = 0.467$ by a substantial factor
varying approximately between 1.3 and 4.0. This also
leads to a bigger (by a factor $\sim (1.5 - 2.0)$) \night\ 
asymmetry \AsymPN\ for the indicated values of 
$sin^22\theta$. Thus, the \core\ asymmetry \AsymPC\ 
at small mixing angles 
is considerably more sensitive to the value of 
$Y_e$ in the Earth core
when the solar $\nu_e$ transitions are into 
sterile neutrino than if the transitions
were of the type 
$\nu_e \rightarrow \nu_{\mu (\tau)}$ \cite{ArticleI}. 
\vspace{-0.3cm}
\section{\daynight\ Effect Related Observables}

\indent  As explained in the preceding Section, 
following the analyzes performed  
in \cite{ArticleI, ArticleII} we consider four possible groups or 
samples of solar neutrino events in the 
Super-Kamiokande experiment depending on their detection time:
\DAY, \night, \core\ and \mantle.
The recoil-\electron\ spectra associated with the four samples are
denoted by $\Ses(\Te)$, where $s = D$, $N$, $C$, $M$, and \Te\ is
the recoil-\electron\ kinetic energy, while for the event rates we will use 
the notation \Res.
The symbols $\SeZr(\Te)$\ and $\ReZr$\ will be used to
denote the recoil-\electron\ spectrum and event rates for massless 
(``conventionally'' behaving) neutrinos, computed using 
the predictions of a given 
reference standard solar model. Obviously,
$\SeZr(\Te)$\ and $\ReZr$\ are the same for the four
event samples of interest.
The spectra $\SeZr(\Te)$, $\Ses(\Te)$\ and  the event rates $\ReZr$, \Res\
considered in the present article 
are one year averaged spectra and 
event rates.

  The spectra $\SeZr(\Te)$\ and $\Ses(\Te)$\ are given by the 
well-known expressions:
\vspace{-0.4cm}
\bec\beq\label{eq:electron:spectra:SSM}
{\displaystyle
     \SeZr(\Te) = \Phi_B
     {\displaystyle \int_{ \Te\left(1 + 
            {{m_e}\over{2 T_e}}\right)} 
}
       d\Enu\, n(E_{\nu})\, \dseedEe
},
\eeq\eec

\noindent
and
\vspace{-0.6cm}
\bec\beq\label{eq:electron:spectra:msw}
    \Ses(\Te) = \Phi_B {\displaystyle \int_{ \Te\left(1 +
                     {{m_e}\over{2 T_e}}\right)} 
                     }
                     d\Enu\, 
                     n(E_{\nu}) \,
 \dseedEe P_\oplus^s\left( \nue \rightarrow \nue \right).
\eeq\eec

\noindent Here $\Phi_B$ is the total \BHt\ neutrino flux, 
\Enu\ is the incoming \BHt\ neutrino energy, $m_e$\ is the electron mass, 
$n(E_{\nu})$ is the normalized to one \BHt\ neutrino spectrum
\cite{Bahcall:etal:1996},
$\PTots(\nue \rightarrow \nue)$\ is the one year averaged solar \nue\ 
survival probability for \DAY, \night, \core, and \mantle\ samples, and
$d \, \sigma_{\nu_e} (\Te,\Enu) / d\,\Te$\ is the
differential $\nue~ - \electron$\ elastic scattering cross section
\cite{Bahcall:Kamionkowski:Sirlin:1995}.
Note the absence in eq. (16) of an analog of 
the neutral current
$\numt - \electron$\ elastic scattering term 
which plays an important role
in the case of $\nu_e \rightarrow \nu_{\mu (\tau)}$ transitions.
For the corresponding energy integrated event rates we have:
\vspace{-0.6cm}
\bec\beq
\begin{array}{lll}
  \ReZr(\TeTh) &=&  {\displaystyle \int_{\TeTh}
} 
d\Te\,\SeZr(\Te),\\
   &&\\
  \Res(\TeTh) &= & {\displaystyle \int_{\TeTh}
 } d\Te \, \Ses(\Te),\\
\end{array}
\eeq\eec

\noindent
where \TeTh\ is the recoil-\electron\ kinetic energy threshold of the \SK\ 
detector. In order to assess the dependence of the magnitude of 
D-N effect in the three event 
samples on \TeTh\ we report here results obtained for two values
of \TeTh: $\TeTh = 5$\ MeV and 7.5 MeV.

  As like in \cite{ArticleI}, we have studied three observables relevant to 
\daynight\ effect which can be measured with the \SK\ detector. 
The first is the distortion of the recoil-\electron\ spectrum due to the 
MSW effect for the four different event samples:
\vspace{-0.8cm}
\bec\beq
N^{s}(T_e) = \frac{\Ses(\Te)}{\SeZr(\Te)}, \mbox{\hspace{1cm}} 
            s = \mbox{$D$, $N$, $C$, $M$}.
\eeq\eec

\noindent
In the absence of the MSW effect, or in the case of energy - independent 
(constant) reduction of the \BHt\ \nue\ flux at $\Enu \ge 5$ MeV, we 
would have $N^{s}(T_e) =~ const$. 
The spectrum ratio $N^{s}(T_e)$ shows the magnitude and 
the shape of the \electron\ spectrum deformations 
(with respect to the standard spectrum)
due to the MSW effect taking place in the Sun only, as well as 
in the Sun and when the solar $\nu_e$ cross the Earth mantle only, 
the Earth core, or the mantle only + the core.  

The second observable we consider is the \daynight\ asymmetry in the 
recoil-\electron\ spectrum for the three solar neutrino event samples, 
associated with the MSW effect in the Earth:
\vspace{-0.8cm}
\bec\beq
     \AsymSs(\Te) = 2 \,\, \frac{\Ses(\Te)-\SeD(\Te)}{\Ses(\Te)+\SeD(\Te)},
      \mbox{\hspace{1cm}}
       s = \mbox{$N$, $C$, $M$}.
\eeq\eec

\noindent The third observable is the energy integrated event rate asymmetry
\footnote{Note the difference between the definitions of the event rate
D-N asymmetry used by us and that employed by the Super-Kamiokande
collaboration, eq. (7).}:
\vspace{-0.8cm}
\bec\beq
     \AsymRs (T_{e,th}) \equiv \AsymRs = 2 \,\, \frac{\Res - \ReD}{\Res+\ReD}, 
      \mbox{\hspace{1cm}}
       s= \mbox{$N$, $C$, $M$}.
\eeq\eec
\noindent The shape of the $^{8}$B neutrino spectrum and the asymmetries
\AsymSs(\Te)\ and  $\AsymRs (T_{e,th})$ 
are solar model independent observables.

   The observation of a nonzero \daynight\ 
asymmetry $\AsymSs(\Te) \neq 0$\ and/or $\AsymRs (T_{e~th}) \neq 0$\   
would be a very strong 
evidence (if not a proof) that 
solar neutrinos undergo MSW transitions. 
If, for instance, nonzero \core\ and \mantle\ (or \night ) asymmetries
will be observed, the magnitudes of these asymmetries can 
indicate, in particular, whether the solar $\nu_e$ 
undergo transitions into active or sterile neutrinos (see further).

  The one year averaged spectrum ratio 
$N^{s}(T_e)$, $s= D$, $N$, $C$, $M$,
and the spectrum and event rate \daynight\ asymmetries $\AsymSs(\Te)$\ and 
$\AsymRs$, $s= N$, $C$, $M$, for the \SK\ detector have been calculated for
39 pairs of values of \dms\ and \SdTvS\ distributed evenly in the 
``conservative'' MSW $\nu_e \rightarrow \nu_s$ transition 
solution regions, eqs. (3) - (4) and (5) - (6). 
These values together with the corresponding results for 
\AsymRNCM\ and for the ratio $\AsymRC / \AsymRN$,
are given in Tables I - VII. 
The predicted event rates for the \DAY, \night, \core\ and \mantle\
samples for the two values of
$Y_e$ in the Earth core considered by us, 
$Y_e(core) = 0.467$ and 0.50, are given 
in Tables I, V, VI and VII 
in units of the rates, calculated in the standard solar model
\cite{BP95} with ``conventionally'' behaving solar 
neutrinos and multiplied by the $^{8}$B neutrino flux factor 
$f_{B} = \Phi_{B} /\Phi^{BP}_{B}$ 
(see, e.g., \cite{KPL96} and \cite{ArticleII})
given in the Tables.
Tables II - IV contain the results for the D-N asymmetries
$\AsymRs$, $s= N$, $C$, $M$, and for the ratio
$\AsymRC / \AsymRN$ for $Y_e(core) = 0.467~and~ 0.50$.
The asymmetries given in Tables II - IV have been
calculated for two values of the threshold 
recoil$-e^{-}$ kinetic energy: 
$T_{e,th} = 5.0~{\rm MeV}~and~ 7.5~{\rm MeV}$.
Some of the results 
obtained for $N^{s}(T_e)$ and $\AsymSs(\Te)$\ 
are presented graphically in Figs. 4.1 - 4.16.  
Finally, Figs. 5a - 5b, Figs. 6a - 6d, and Figs. 7a - 7b  
contain plots of contours 
in the $\dms$\ - $\SdTvS$\ plane,
corresponding respectively to fixed values of the event rate asymmetries 
\AsymRN, \AsymRC, and \AsymRM\ (iso - (\daynight) asymmetry contours)
in the region of $\SdTvS \geq 10^{-4}$, 
$10^{-7}~{\rm eV^2} \leq \dms \leq 10^{-4}~{\rm eV^2}$. 
The iso-(D-N) asymmetry contour
plots are derived for electron number fraction in the core 
$Y_e(core) = 0.467$ (Figs. 5a, 5b, 6a and 6c) and 
$Y_e(core) = 0.50$ (Figs. 5a, 5b, 6b and 6d), and for 
$T_{e,th} = 5.0~{\rm MeV}$ (Figs. 5a, 6a, 6b, and 7a) 
and  $T_{e,th} = 7.5~{\rm MeV}$ (Figs. 5b, 6c, 6d and 7b).
\vspace{-0.3cm}
\section{The Earth Matter Effect and the Recoil$-e^{-}$ Spectrum}

\indent   We shall discuss first the relation between the structures
(``peaks'') present in the probability asymmetries \AsymPs\ 
(due to the Earth matter effect) and 
the structures in the corresponding spectrum D-N asymmetries \AsymSs(\Te).
For a given $\sin^22\theta$ the probabilities 
\APeTws\ and \PTots\ and the asymmetries \AsymPs\ are functions of
$E_{\nu}/\Delta m^2$. If $\Delta m^2$ is given, a structure
or ``peak'' in  $\AsymPs (E_{\nu}/\Delta m^2)$ taking place at
$E_{\nu}/\Delta m^2 = (E_{\nu}/\Delta m^2)_{p}$,
would correspond to a structure  
in the solar neutrino 
distorted spectrum at $E_{\nu p} = \Delta m^2 (E_{\nu}/\Delta m^2)_{p}$.
Qualitatively the effect of a ``peak'' 
in \AsymPs\ will be maximal if the 
corresponding $E_{\nu p}$ is near the maximum 
in the neutrino spectrum, which for $^{8}$B 
neutrinos is located at approximately 6.5 MeV.  
A necessary condition for the structure to appear in the 
solar neutrino 
spectrum obviously is 
$E_{\nu}^{min} \leq E_{\nu p} \leq E_{\nu}^{max}$,
where $E_{\nu}^{max}$  ($E_{\nu}^{min}$) is the maximal (minimal)
solar $\nu_e$ energy to which a given detector is sensitive.
In the case of the Super-Kamiokande experiment one has:
$E_{\nu}^{max} \cong 14.4~{\rm MeV}$, which is the maximal energy of
the $^{8}$B neutrino flux, and $E_{\nu}^{min} \cong 5.26~{\rm MeV}~ 
(7.76~{\rm MeV})$ if the threshold recoil-$e^{-}$ kinetic energy is
5 MeV (7.5 MeV). Further, 
in the indicated energy intervals, 
the cross-section $d\sigma_{\nu_e}(\Te,\Enu)/d\Te$,
which enters into eq. (18) and is proportional to 
the probability that a solar neutrino 
having an energy $E_{\nu}$ will produce a recoil electron 
with a kinetic energy $T_e$, is practically $T_e-$independent 
and depends linearly on $E_{\nu}$. 
This can lead, in particular, to a reduction in magnitude or 
disappearance (``filling'') of the ``peaks'' present in the
probability asymmetries \AsymPs (\EDms)\ when one 
calculates the \daynight\ asymmetries in the recoil-e$^{-}$ spectrum, 
\AsymSs(\Te). To illustrate 
this point consider two ``peaks'' in \AsymPs\ located at 
$E_{\nu}/\Delta m^2 = (E_{\nu}/\Delta m^2)_{1}$ and
at $E_{\nu}/\Delta m^2 = (E_{\nu}/\Delta m^2)_{2}$,
with  $(E_{\nu}/\Delta m^2)_{2} < (\EDms)_{1}$.
Let us assume that 
$\AsymPs((\EDms)_2) > 0$ and $\AsymPs((\EDms)_1) < 0$.
If for a chosen fixed $\Delta m^2$ one has, say, 
$E_2 < 14~ {\rm MeV}$ and 
$15~{\rm MeV} < E_1 $, where 
$E_{2,1} = \dms (\EDms)_{2,1}$, 
the asymmetry $\AsymSs(\Te)$\ will also exhibit a ``peak'' 
at \Te\ located in the interval determined by the inequality: 
$0.5\Te (1 + \sqrt{1 + 2m_e/\Te)}) < E_2$.
However, if \dms\ is such that $E_{1,2} < 14~{\rm MeV}$,  
the integration over $E_{\nu}$  in eq. (18) can lead to 
a partial (or complete compensation)
between the contributions in \AsymSs(\Te)\ from the regions of 
the positive and the negative ``peaks'' 
in \AsymPs(\EDms) when \Te\ lies in the 
indicated interval.
As a consequence, in such intervals 
$|\AsymSs(\Te)|$ can be considerably smaller than 
the corresponding $|\AsymPs(\EDms)|$. It should be obvious from the 
above and the preceding discussions that 
the asymmetry \AsymSs(\Te)\ can exhibit a 
strong dependence on the value of
\dms, and that, in general, one can expect the extrema 
and the bulk of $\AsymSs(\Te)$ to be located
at relatively large values of \Te, $\Te~\gtap (8 - 9)~{\rm MeV}$.
Let us add that due to the differences between
the \EDms-dependence of the probabilities \PTots\ corresponding to 
$\nu_e \rightarrow \nu_s$ and to
$\nu_e \rightarrow \nu_{\mu (\tau)}$ transitions, 
discussed in Section 3, at small mixing angles
these considerations are much more relevant 
for the $\nu_e \rightarrow \nu_s$ case
than for the $\nu_e \rightarrow \nu_{\mu (\tau)}$ one.

 The above remarks are well illustrated by Figs. 4.1 - 4.16, in which  
we show the recoil$-e^{-}$ spectrum distortions
$N^{s}(T_e)$ (upper frames) and \daynight\ asymmetries \AsymSs(\Te)\ 
(lower frames) for selected representative sets of 
the parameters \dms\ and $\sin^22\theta_V$, as 
functions of the recoil-e$^{-}$ kinetic energy \Te.
The {\em short - dashed}, {\em solid} and {\em dotted} lines correspond to 
spectrum distortions or \daynight\ asymmetries for the \night, \core\ and 
\mantle\ samples respectively, while the {\em long - dashed} lines correspond 
to the spectrum distortion for the \DAY\ sample.
Figures 4.1 - 4.12 and figures 4.13 - 4.16 have been obtained for
$\Ye (core) = 0.467$ and for $\Ye (core) = 0.50$, respectively.

  At small mixing angles, $\SdTvS < 0.014$, 
the spectrum \daynight\ asymmetry for the \night\ sample,
\AsymSN(\Te), is very small, $|\AsymSN(\Te)|~\ltap~ 0.01$ for
$5~{\rm MeV} \leq T_e \leq 14~{\rm MeV}$, and seems hardly
observable with the Super-Kamiokande detector. This conclusion is valid
both for $Y_e(core) = 0.467$ and $Y_e(core) = 0.50$. For 
$\sin^22\theta \geq 0.014$ the spectrum deformations due to the
Earth matter effect and the asymmetry \AsymSN(\Te)\
increase with the increasing of
$\sin^22\theta$ and become nonnegligible at large mixing angles.
For $0.4 ~\ltap ~\sin^22\theta ~\ltap~ 0.5$, for instance,
the asymmetry \AsymSN(\Te)\ is typically bigger than
(15 - 20)\% and for $Y_e(core) = 0.467$ ($Y_e(core) = 0.50$) it can
be as large as 40\% (50\%) in the interval of values of 
$T_e = (5 - 14)~{\rm MeV}$ of interest.
 
    In the case of the \core\ sample, 
the effect of the core enhancement is evident both in 
the $e^{-}-$spectrum distortions and in the corresponding 
\daynight\ spectrum 
asymmetry \AsymSC(\Te), but at small 
mixing angles both are hardly detectable with the 
Super-Kamiokande detector. Because of 
the effect of ``spectrum filling'' discussed above,
 the largest value of $|\AsymSC(\Te)|$
is typically smaller
than the largest value of $|\AsymPC(\EDms)|$.
For $\YeCore = 0.467$ and at
$\SdTvS \leq 0.007$\ the negative \daynight\ 
effect is the only relevant feature and 
for $\SdTvS > 0.001$\ the negative spectrum asymmetry 
can be larger in absolute value than $1\%$ and can reach a maximum of 3\%, 
provided the \dms\ value does not ``push'' the region
of the negative ``peak'' in \AsymPC\ above the 
$^{8}$B neutrino spectrum upper end. The latter typically occurs for
$\dms ~\gtap~ (7 - 8)\times 10^{-6}~{\rm eV^2}$.
Due to the ``spectrum filling'', 
positive values of $\AsymSC(\Te) \gtap~1\%$ do not  
appear in the interval $\Te = (5 - 14)~{\rm MeV}$ 
as long as $\sin^22\theta_V < 0.008$. 
At $\SdTvS = 0.009$, for instance, \AsymSC(\Te)\ takes
both positive and negative values going through zero in the 
indicated interval if $\dms = 4\times 10^{-6}$ eV$^2$ 
(Figs. 4.6 and 4.13),
while for $\dms = 5\times 10^{-6}$ eV$^2$\ (Figs. 4.7 and 4.14) 
the ``filling'' is less relevant 
since the negative `` peak'' in \AsymPC\ is
located (in $E_{\nu}$) above the upper limit of 
the $^{8}$B neutrino spectrum.
For $\SdTvS~\cong 0.014$ we have $\AsymSC(\Te) > 0$ 
and the spectrum asymmetry can be as large as 16\% for $Y_e(core) = 0.467$.

   At large mixing angles (Figs. 4.10 - 4.12), the presence of a 
negative \daynight\ effect in \AsymPC(\EDms)\ (Figs. 3.11 - 3.12) 
is reflected in \AsymSC(\Te). However, together with
the dominant positive ``peak'' in \AsymPC(\EDms), it is located 
(in $E_{\nu}$) beyond the $^{8}$B neutrino spectrum upper limit
if $\dms > 8.5\times 10^{-6}~{\rm eV^2}$,
and therefore both have little effect on any of the
D-N effect observables  
for $\dms > 8.5\times 10^{-6}~{\rm eV^2}$
(see, e.g., Fig. 4.10).
The ``spectrum filling'' effect is evident at $\SdTvS = 0.4$, for instance, 
if we compare the two \AsymSC(\Te)\ plots, Figs. 4.10 and 4.11,
obtained for $\dms = 6\times 10^{-6}$ eV$^2$ and for 
$\dms = 8 \times 10^{-6}$ eV$^2$. 
At $\dms = 6\times 10^{-6}$ eV$^2$, the spectrum asymmetry
\AsymSC(\Te)\ changes from (+50\%) at $\Te \cong 7~{\rm MeV}$ to
(-50\%) at $\Te \cong 12~{\rm MeV}$.
For $\SdTvS = 0.5$, the asymmetry \AsymSC(\Te)\ is maximal
at $\dms \cong 8\times 10^{-6}$ eV$^2$, reaching approximately the
value of +70\% at $\Te \cong 11.5~{\rm MeV}$.
The \night\ and \mantle\ spectrum asymmetries, \AsymSN(\Te)\ and 
\AsymSM(\Te), are positive at large mixing angles,
having values between 15\% and 45\%.

   Changing the value of $\YeCore$
from 0.467 to 0.50 leads to small (insignificant) changes of the spectrum
asymmetries \AsymSC(\Te)\ and \AsymSN(\Te)\ as long as 
$\sin^22\theta~ \ltap ~0.007$. 
However, for $\SdTvS \geq 0.008$, the asymmetry $|\AsymSC(\Te)|$   
increases by a factor which can be as large as 2.
For $\SdTvS = 0.5$ and, e.g., 
$\dms = 8\times 10^{-6}$ eV$^2$,  
\AsymSC(\Te)\ reaches 85\%, so that a 7\%
increase in $\YeCore$ produces nearly a 20\% increase 
in the maximal value of \AsymSC(\Te).
\vspace{-0.3cm}
\section{Event Rate \daynight\ Asymmetries}

\indent Our results for the magnitudes of the event rate D-N asymmetries 
in the \night, \core\ and \mantle\ samples are collected in Tables II - IV
and in Figs. 5a - 5b, 6a - 6d and 7a - 7b. In what follows we will 
comment on these results.

  We begin with a discussion of the \night\ and \mantle\ sample asymmetries
to be expected for \dms\ and $\sin^22\theta$ from the 
``conservative'' $\nu_e \rightarrow \nu_s$ solution regions.
For $\sin^22\theta ~\leq 0.010$ the \night\ asymmetry is never bigger than
approximately 1.25\%: $|\AsymRN| ~\ltap~1.25\%$. 
Actually, the value of 1.25\% is reached  for $\Ye (core) = 0.50$ and 
$T_{e,th} = 7.5~{\rm MeV}$ at $\sin^22\theta = 0.010$ (Table IV). 
For $T_{e,th} = 5.0~{\rm MeV}$ and $\Ye (core) = 0.467~(0.50)$, 
we have in the indicated small mixing angle region (Table II): 
$|\AsymRN| ~\ltap~0.70\%~(1.17\%)$, the maximal values corresponding 
again to $\sin^22\theta = 0.010$. The asymmetry in the \mantle\ sample,
\AsymRM, is typically smaller in absolute value than the \night\ sample 
asymmetry, $|\AsymRN|$, by a factor of (1.5 - 2.0).
 
  At $\sin^22\theta \cong 0.014$ 
the asymmetry \AsymRN\ is considerably 
larger: for $\Ye (core) = 0.467$ and
$T_{e,th} = 5.0~(7.5)~{\rm MeV}$,
it varies approximately between
1.8\% (2.0\%) and 3.0\% (3.7\%) 
depending on the solution value of \dms. The change of the threshold
energy from 5.0 MeV to 7.5 MeV leads to an increase 
of \AsymRN. The increase is much more dramatic - approximately 
by a factor of (1.5 - 2.0), if $\Ye (core) = 0.50$ instead of 
0.467 (Table II).
The \mantle\ asymmetry continues to be 1.5 to 2.0 times 
smaller than the \night\ asymmetry and changes insignificantly
when $T_{e,th} = 5.0~{\rm MeV}$ is replaced by
$T_{e,th} = 7.5~{\rm MeV}$. 

  It should be noted that as a result of the differences
between the Earth matter effect 
in the cases of the $\nu_e \rightarrow \nu_s$ and the
$\nu_e \rightarrow \nu_{\mu(\tau)}$
solutions of the solar neutrino problem discussed in detail
in Sections 2, 3 and 5, at small mixing angles,
$0.006 ~\ltap \sin^22\theta ~\ltap 0.01$,
the \night\ and the \mantle\ D-N asymmetries under study
are substantially smaller than the asymmetries 
predicted for the $\nu_e \rightarrow \nu_{\mu(\tau)}$
solution \cite{ArticleI}. For $\sin^22\theta = 0.008~(0.01)$,
for instance, and for $\Ye (core) = 0.467$ and
$T_{e,th} = 5.0~{\rm MeV}$,
the \night\ sample asymmetry  
corresponding to the $\nu_e \rightarrow \nu_{\mu(\tau)}$ 
solution, $\AsymRN (active)$, exceeds the asymmetry $|\AsymRN (sterile)|$   
in the case of the $\nu_e \rightarrow \nu_{s}$ solution 
by a factor which, depending on the chosen values of \dms\ from the 
corresponding solution regions, varies between 
\footnote{For the same \dms\ values from the overlapping region of the two 
solutions the factor ranges between approximately 4 and 15.}
approximately 4 and 25
(see Table II in \cite{ArticleI} and Table II in this article).  
The analogous factor 
for the \mantle\ asymmetry 
is smaller: it varies typically between 2 and 9.
  
  At large mixing angles, $\sin^22\theta \cong (0.40 - 0.50)$,
the \night\ and the \mantle\ asymmetries are of the same order,
both exceeding 18\% and having maximal values of $\sim 40\%$.
Both asymmetries \AsymRN\ and \AsymRM\ increase by 
10\% to 20\%, depending on the value of 
\dms\ from the solution region,
when $\Ye (core) = 0.467$ is changed to 
$\Ye (core) = 0.50$ and/or when 
$T_{e,th} = 7.5~{\rm MeV}$ is used instead of 
$T_{e,th} = 5.0~{\rm MeV}$. Both \AsymRN\ and 
\AsymRM\ have their smaller values in 
the considered $\sin^22\theta$ interval
for $\Ye (core) = 0.467$ and $T_{e,th} = 5.0~{\rm MeV}$.

  Adding the systematic and the statistical errors in eq. (7) 
in quadratures, we find that at 95\% C.L. the Super-Kamiokande result
on the D-N asymmetry in the \night\ sample implies:
$\AsymRN~\leq 17.4\%$. This rules out (at the indicated confidence level)
the ``conservative'' adiabatic region (eqs. (5) - (6)) of 
the $\nu_e \rightarrow \nu_s$ transition solution.
The large mixing angle $\nu_e \rightarrow \nu_s$ adiabatic solution  
was possible only for rather large values of the
$^{8}$B neutrino flux \cite{KPL96}. 

    Consider next the \core\ asymmetry, \AsymRC. 
At small mixing angles and for values of \dms\ from
the ``conservative'' solution region, 
$|\AsymRC|$ is bigger (due to the core enhancement) 
than $|\AsymRN|$ by a factor of approximately (3 - 4)
if $\dms ~< 5\times 10^{-6}~{\rm eV^2}$, and by a factor
of $\sim (1.5 - 2.5)$ for $\dms ~\gtap 5\times 10^{-6}~{\rm eV^2}$
(Table II).
In the solution region where
$0.0012 ~\ltap ~\sin^22\theta ~\ltap ~0.008$ and 
$\dms ~\ltap ~4\times 10^{-6}~{\rm eV^2}$,
one has $(-2\%)~\ltap ~\AsymRC ~\ltap ~(-1\%)$. For 
$\sin^22\theta ~\gtap 0.009$ 
and $\dms ~\ltap~ 4.4\times 10^{-6}~{\rm eV^2}$ 
the asymmetry is positive and $\AsymRC \geq 1\%$.
The asymmetry \AsymRC\ has a minimum
in the interval $0.008~\ltap~\sin^22\theta ~\ltap ~0.03$ 
at $\dms \cong 6.0\times 10^{-6}~{\rm eV^2}$ and for this value
of \dms\ we have $\AsymRC \geq 1\%$ only when $\sin^22\theta \geq 0.012$.
Changing the threshold energy $T_{e,th}$ from 5 MeV to 7.5 MeV 
typically increases $|\AsymRC |$ 
for $\sin^22\theta ~\ltap~ 0.03$
by a factor of $\sim (1.2 - 1.5)$ 
if $\dms ~\ltap ~5\times 10^{-6}~{\rm eV^2}$,
decreases somewhat $|\AsymRC |$ if 
$\dms ~\cong ~(6 - 7)\times 10^{-6}~{\rm eV^2}$, and 
practically does not change the asymmetry
when $\dms ~\sim 5.5\times 10^{-6}~{\rm eV^2}$ (Table III).
In the region
$0.0012 ~\ltap ~\sin^22\theta ~\ltap ~0.008$,
$\dms ~\ltap ~4\times 10^{-6}~{\rm eV^2}$, for instance, for 
$T_{e,th} = 7.5~$MeV we get 
$(-3\%)~\ltap ~\AsymRC ~\ltap ~(-1\%)$.
The asymmetry \AsymRC, however, is rather 
sensitive to the value of
\Ye (core). The dependence of \AsymRC\ on \Ye (core)
is particularly strong in the solution interval
$0.0075 ~\ltap ~\sin^22\theta ~\ltap ~0.030$, where a change
of the value of \Ye(core)\ from 0.467 to 0.50 leads to an
increase of \AsymRC\ by a factor of $\sim ( 2 - 4)$.

  The \core\ sample asymmetry in the case of the 
$\nu_e \rightarrow \nu_{s}$ solution under study, \AsymRC (sterile), 
is substantially
smaller than the \core\ asymmetry, \AsymRC (active), predicted for the 
$\nu_e \rightarrow \nu_{\mu(\tau)}$ transition solution
(see Table II in \cite{ArticleI} and Table II in this article):
the ratio of the absolute values of the two \core\ asymmetries
$|\AsymRC (active)/\AsymRC (sterile)|$
in the small mixing angle
region $0.006 ~\ltap ~\sin^22\theta ~\ltap ~0.014$ where
$\AsymRC ({\rm active})~\geq 1\%$, 
is always greater than 3.5, can be as large as 40
(for, e.g., $\dms = 7\times 10^{-6}~{\rm eV^2}$
and $\sin^22\theta = 0.008$) and is typically 
\footnote{As our results show, the naive suggestion
that $|\AsymRs (sterile)| > |\AsymRs (active)|$, made e.g. 
in ref. \cite{QLAS97} and the discussion based on it, are incorrect.}
between 5 and 10.

   The specific features of the three energy-integrated 
event rate D-N asymmetries,
\AsymRN, \AsymRC\ and \AsymRM, 
discussed above can also be seen on the iso-(D-N) asymmetry 
contour plots derived in the region 
$10^{-7}~{\rm eV^2} \leq \dms \leq 10^{-4}~{\rm eV^2}$,
$10^{-4} \leq \sin^22\theta \leq 1.0$ and
shown in Figs. 5a - 5b, 6a - 6d and 7a - 7b~~
\footnote{Let us note that the most interesting 
feature in \AsymRC\ at large mixing angles 
is the negative \daynight\ effect producing a 
``hole'', i.e., a deep minimum, in $\AsymRC$\  
at $\SdTvS \approx 0.5$, $\dms \approx 3.5\times 10^{-6}$ eV$^2$.
The ``hole'' is clearly seen in Figs. 6a - 6b (Figs. 6c - 6d)
where for graphical reasons only the iso-asymmetry contour 
lines for $\AsymRC \geq -0.02~(-0.03)$ are 
displayed in the region of the ``hole''. 
The value of the asymmetry reached at the 
``bottom'' of the ``hole'' for 
$\Ye (core) = 0.467$\ is 
$\AsymRC = -49\%~(-51\%) $\ for $T_{e,th} = 5.0 ~(7.5)$\ \MeV,
while if $\YeCore = 0.5$\ the minimum value reads $-51\%~(-91\%)$.}.

  It follows from the above results that probing experimentally
the region of the SMA solution via the D-N effect 
with the Super-Kamiokande detector will require
a much higher precision in the case of solar neutrino
transitions into sterile neutrino than if the solar $\nu_e$
transitions were into an active neutrino. 

\vspace{-0.3cm}

\def\AsymRCO{\mbox{$A_{D-N,O}^C$}}     
\def\AsymRMO{\mbox{$A_{D-N,O}^M$}}     

\section{D-N Asymmetry Measurements: Discriminating Between 
the $\nue \rightarrow \numt$\ and 
        the $\nue \rightarrow \nus$\ Transition Solutions}
 
\indent  If two-neutrino MSW transitions of the solar neutrinos are at
the origin of the solar neutrino problem, 
it will be of fundamental importance to
determine whether the transitions are into active or into 
sterile neutrinos, or into a mixture of active and sterile neutrinos. 
The solar $\nue \rightarrow \nus$\ transitions produce 
(for given \dms\ and $\sin^22\theta$)
larger distortions of the recoil-e$^{-}$ spectrum
which is being measured in the Super-Kamiokande experiment, 
than the $\nue \rightarrow \numt$\ transitions \cite{KPL96}
(see also \cite{KWROSEN96}). 
The two MSW solutions of the solar neutrino 
problem imply very different values of
the ratio of the charged current (CC) and the neutral current (NC)
event rates, $R_{CC/NC}$, to be measured with the SNO detector
\cite{KPUNPUB96}. An unambiguous proof that the solar neutrinos
undergo transitions into sterile neutrino would be
the observation of a nonzero D-N asymmetry in the NC
signal due to the solar neutrinos in the SNO detector \cite{KPL96}:
the indicated asymmetry is zero in the case of
$\nue \rightarrow \numt$\ transitions.  Other possible tests
of the hypothesis of $\nue \rightarrow \nus$\ conversion of
solar neutrinos which can be performed 
utilizing the Super-Kamiokande and SNO data
were considered in ref. \cite{BilGiunt95}.

       If the solar neutrinos undergo two-neutrino matter-enhanced 
transitions generated by the existence of
nonzero neutrino masses and mixing, 
the measurement of the D-N asymmetries in 
the \core\ and the \mantle\  samples of events, 
which are two independent observables,
can also give information about the type of the transitions:
$\nue \rightarrow \nus$\ or $\nue \rightarrow \numt$.
In the brief qualitative discussion which follows we  
neglect the uncertainties in the Earth core density and chemical 
composition, consider the case of $\TeTh = 5$ MeV and analyze
the small mixing angle solutions only. 
A complete quantitative study of the indicated
possibility is beyond the scope of the present
paper and will be given elsewhere
\cite{Maris:Petcov:1998}.

  The mean event rate data from the different solar 
neutrino experiments determine the 
$\nue \rightarrow \nus$\ and $\nue \rightarrow \numt$
solution regions in the $\dms - \sin^22\theta_{V}$ plane.
With the improvement of the data the regions will diminish
in size. The  values of the two D-N effect asymmetries 
\AsymRC\  and \AsymRM, derived from the data, must correspond
to one and the same set (or range) of values $\dms$ and  $\sin^22\theta_{V}$.
Suppose, for instance, that the measurement of the \mantle\ sample
asymmetry gives the upper limit: $\AsymRM < 2.5\%$.
For the small mixing angle solution values of
\dms\ and  $\sin^22\theta_{V}$, which are compatible with
the above limit, the asymmetries 
$\AsymRC (active)$ and $\AsymRC (sterile)$ corresponding to
the $\nue \rightarrow \nus$\ and $\nue \rightarrow \numt$
transitions are also limited from above:
$\AsymRC (active) \leq 30\%$ and $\AsymRC (sterile) \leq 20\%$
(see Figs. 3b and 3c in ref. \cite{ArticleII} and Figs. 6a and 7a).
An observation of a \core\ sample asymmetry which is definitely greater
than 20\% would imply that the solar $\nu_e$ transitions are of the type
$\nue \rightarrow \numt$, while if the measured value of
\AsymRC\ is smaller than 20\%, no unambiguous conclusion about the type of
the transition can be drawn.
Similarly, an experimental upper limit on \AsymRM\ of 1\% and
a measured value of the \core\ sample asymmetry
bigger than 10\% would mean that the solar neutrinos undergo 
transitions into an active neutrino since in this case
$\AsymRC (sterile) \leq 10\%$. It should be obvious that
if only upper limits on the \core\ and the \mantle\ sample asymmetries
would be obtained from the data, no definite conclusion concerning 
the type of the solar neutrino transitions could be made.

\vspace{-0.3cm}
\section{Conclusions}

\indent  In the present article we have performed a 
rather detailed quantitative study
of the D-N effect for the Super-Kamiokande detector 
for the solution of the solar neutrino problem
involving two-neutrino matter-enhanced transitions of the 
solar neutrinos into a sterile neutrino,
$\nu_e \rightarrow \nu_s$. The one year average D-N asymmetry, \AsymRs,
has been calculated (using the high precision methods developed
in refs. \cite{ArticleI,ArticleII}) for three samples of events,
\mantle\ (M) , \core\ (C) and \night\ (N), produced
respectively by the solar neutrinos crossing the Earth mantle only,
the Earth core, and by the only mantle crossing + the core 
crossing neutrinos (the full night sample). 
The asymmetry calculations require the knowledge of the
one year averaged spectrum of the recoil electrons and  
energy-integrated even rate, produced by the solar neutrinos 
during the day (the \DAY\ sample).
Results for the D-N asymmetry in the
recoil-e$^{-}$ spectrum for the same three samples of events, \AsymSs (\Te),
s=N,C,M, have also been obtained.
The asymmetries have been calculated for a large representative set of
values of the neutrino transition parameters \dms\ and
$\sin^22\theta$ from the ``conservative'' 
$\nu_e \rightarrow \nu_s$ transition solution regions (eqs. (3) - (6)),
derived by taking into account the possible uncertainties
in the predictions for the $^{8}$B and $^{7}$Be neutrino fluxes. 
We have investigated the dependence of the
three D-N asymmetries studied, on the recoil-e$^{-}$ kinetic energy 
threshold $T_{e,th}$, which can be varied in the 
Super-Kamiokande experiment, by performing calculations of
all the indicated D-N asymmetries 
for $T_{e,th} = 5.0~$ MeV and $T_{e,th} = 7.5~$ MeV.      
The effect of the estimated uncertainties
in the knowledge of the bulk matter density and 
the chemical composition of the Earth core \cite{Stacey:1977,PREM81,CORE} 
on the predictions for the D-N asymmetries, has been studied as well 
by deriving results for \AsymSs (\Te)\ and \AsymRs\
both for the standard value of the
electron number fraction in the core $\Ye (core) = 0.467$ and for 
the estimated conservative upper limit on $\Ye (core)$,
$\Ye (core) = 0.50$ (see \cite{CORE} and
\cite{ArticleI}). Iso-(D-N) asymmetry contour plots
for the \night, \core\ and \mantle\ samples of events
in the region $10^{-7}~{\rm eV^2} ~\leq ~\dms~\leq 10^{-4}~{\rm eV^2}$,
$10^{-4}~\leq~ \sin^22\theta_V ~\leq~ 1$, have been obtained  
for $T_{e,th} = 5.0~$ MeV and $T_{e,th} = 7.5~$ MeV, and 
for the \night\ and \core\ samples - for
$\Ye (core) = 0.467$ and 
$\Ye (core) = 0.50$. The main results of this study are collected 
in Tables I - VII and are shown graphically in Figs. 4 - 7. 
  
  We have found that, as like in the case of the
$\nue \rightarrow \numt$\ solution, the division of the
data collected at night into a \core\ and \mantle\ samples
is a rather effective method of enhancing the D-N asymmetry
at small mixing angles, 
$0.001~\ltap~ \sin^22\theta_V ~\ltap~ 0.03$:
the asymmetry in the \core\ sample $|\AsymRC|$ 
is larger than the asymmetry in the 
\night\ sample $|\AsymRN|$ typically by a factor of (3 - 4) if
$\dms ~< 5\times 10^{-6}~{\rm eV^2}$,
and by a factor of $\sim (1.5 - 2.5)$ for 
$5\times 10^{-6}~{\rm eV^2}~ \ltap ~\dms ~\ltap 8\times 10^{-6}~{\rm eV^2}$
(Table II). However, the enhancement 
is not as strong as in the case of the $\nue \rightarrow \numt$\ 
transition solution \cite{ArticleI}. Moreover, in the interesting region
$0.005~\ltap~ \sin^22\theta_V ~\ltap~ 0.014$
the D-N asymmetries 
in the \core\ and \night\ samples 
found for the $\nue \rightarrow \nus$\ solution, $|\AsymRC (sterile)|$ and
$|\AsymRN (sterile)|$,
are substantially smaller - 
at least by a factor of 4 and typically
by a factor of 5 to 10, than the asymmetries corresponding
to the $\nue \rightarrow \numt$\ solution, \AsymRC (active) and
\AsymRN (active). Similar conclusion is valid for 
the \mantle\ sample asymmetries.
This remarkable difference in the magnitudes of the asymmetries
$|\AsymRs (sterile)|$ and $|\AsymRs (active)|$ in the corresponding
small mixing angle 
solution regions is a consequence of the different roles the neutron
number density distribution in the Earth $n_{n}(r)$ plays in the
solar neutrino transitions in the two cases:
the $\nue \rightarrow \numt$\ transitions, as is well-known,
depend only on the electron number density distribution, $n_{e}(r)$, while
the  $\nue \rightarrow \nus$\ transitions
depend on the difference    
($n_{e}(r) - 0.5~n_{n}(r)$). In the Sun one has \cite{BP95}
$0.5~n_{n}(r) \ll n_{e}(r)$ and $n_{n}(r)$ influences little the
$\nue \rightarrow \nus$\ transitions. In contrast, due to the neutrality and
approximate isotopic symmetry of the Earth matter,
one has in the Earth: $n_{e}(r) - 0.5~n_{n}(r) \cong 0.5~n_{e}(r)$.
This difference between  the number density distributions
$n_{e}(r)$ and ($n_{e}(r) - 0.5~n_{n}(r)$)
in the Earth is at the origin of the dramatic difference 
between the magnitudes of the D-N asymmetries
corresponding to the small mixing angle 
$\nue \rightarrow \nus$\ and 
$\nue \rightarrow \numt$\ transition solutions discussed above.
Correspondingly, it leads to a shift towards smaller (by a factor of 
$\sim 2$) values of \dms\ and larger values of \SdTvS\ of the
iso - \daynight\ asymmetry contours in the
$\dms - \sin^22\theta_V$ plane corresponding to the 
$\nue \rightarrow \nus$\ solution with respect to the 
analogous contours for the $\nue \rightarrow \numt$\ solution
(compare Figs. 3a - 3c in \cite{ArticleI} with Figs. 5a, 6a and 7a).

     At small mixing angles even the \core\ asymmetry corresponding to the
$\nue \rightarrow \nus$\ solution is rather small (Table II, Fig. 6a):
for $0.0012~\ltap~ \sin^22\theta_V ~\ltap~ 0.008$ and
$2.8\times 10^{-6}~{\rm eV^2}~ \ltap ~\dms ~\ltap 4\times 10^{-6}~{\rm eV^2}$
we find $(-2\%)~\ltap ~\AsymRC (sterile)~\ltap ~(-1\%)$.
For other values of \dms\
from the small mixing angle ``conservative'' solution region
$0.001~ \ltap~\sin^22\theta_V ~\ltap ~0.009$ one obtains
$|\AsymRC (sterile)| \leq 1\%$.
We have $\AsymRC (sterile)~\gtap ~1\%$ in the solution region
$\sin^22\theta_V ~\gtap~ 0.009$ and
$3.0\times 10^{-6}~{\rm eV^2}~ \ltap 
~\dms ~\ltap ~4.4\times 10^{-6}~{\rm eV^2}$.
In addition, \AsymRC (sterile)\ has a minimum
in the interval $0.008~\ltap~\sin^22\theta_V ~\ltap ~0.03$ 
at $\dms \cong 6.0\times 10^{-6}~{\rm eV^2}$ and for this value
of \dms\ one has $\AsymRC \geq 1\%$ only when 
$\sin^22\theta_V \geq 0.012$.
The \night\ and \mantle\ asymmetries are larger than 1\% in absolute value
only if $\sin^22\theta > 0.010$ (Table II, Figs. 5a and 7a). 

  Replacing the threshold energy $T_{e,th} = 5~$MeV with 7.5 MeV 
can, depending on the SMA solution value of \dms,
increase $|\AsymRC|$ (by a factor $\sim (1.2 - 1.5)$),
decrease it somewhat or leave the asymmetry practically the same;
it changes little the magnitudes of \AsymRN\ and \AsymRM
(Tables III and IV, Figs. 5b, 6c and 7b).

  The asymmetries \AsymRC\ and \AsymRN, however, are rather 
sensitive to the value of
\Ye (core) (Tables II - IV and Figs. 5a, 5b and 6a - 6d). 
The dependence of \AsymRC\ and \AsymRN\ on \Ye (core)
is particularly strong in the ``conservative'' solution interval
$0.0075 ~\ltap ~\sin^22\theta_V ~ < ~0.030$, where a change
of the value of \Ye (core)\ from 0.467 to 0.50 leads to an
increase of $|\AsymRC|$ and $|\AsymRN|$ by factors of $\sim ( 2 - 4)$.

     The predicted D-N asymmetries in the recoil-e$^{-}$ spectrum 
for the three samples of events are small 
in the SMA solution region (Figs. 4.1 - 4.16). 
The spectrum asymmetry for the \night\ sample, for instance,
at $\SdTvS < 0.014$ satisfies
$|\AsymSN(\Te)|~\ltap~ 1\%$ for
$5~{\rm MeV} \leq T_e \leq 14~{\rm MeV}$, and is hardly
observable with the Super-Kamiokande detector. This conclusion is valid
both for $Y_e(core) = 0.467$ and $Y_e(core) = 0.50$.
Analogous results are valid for the \core\ sample spectrum asymmetry
\AsymSC (\Te): one has $|\AsymSC(\Te)| \geq 4\%$ only if
$\SdTvS \geq 0.01$; at $\SdTvS \cong 0.014$ the asymmetry 
\AsymSC(\Te) reaches 16\%.

     The upper limit on the D-N asymmetry \AsymRN\ following from the
Super-Kamiokande data (eq. (7)) rules out (at 95\% C.L.) the
``conservative'' large mixing angle (adiabatic) solution  
possible in the case of solar $\nue \rightarrow \nus$\ transitions for   
unrealistically large values of the $^{8}$B neutrino flux \cite{KPL96}. 

    A qualitative analysis performed by us indicates that the 
measurement of the \core\ and \mantle\
sample asymmetries, which are independent observables, can help
to discriminate between the  
$\nue \rightarrow \numt$\ and the $\nue \rightarrow \nus$\
transition solutions of the solar neutrino problem.
 
   The results obtained in the present study 
suggest that it will be difficult to probe
the small mixing angle nonadiabatic  
$\nue \rightarrow \nus$\ transition solution of the solar neutrino
problem at $\SdTvS ~\ltap~ 0.01$
by performing high precision measurements of the 
event rate and the recoil-e$^{-}$ spectrum D-N asymmetries
with the Super-Kamiokande detector.
The precision required to test the indicated solution region
exceeds, for most values of the parameters 
\dms\ and $\SdTvS$ from the region,
the precision in the D-N asymmetry measurements
which is planned to be achieved in the Super-Kamiokande experiment. 

\section*{Acknowledgments}

We are indebted to the ICARUS group of the University of Pavia  
and INFN, Sezione di Pavia, and especially to Prof. E. Calligarich, 
for allowing the use of their computing facilities for the present study.
M.M. wishes to thank Prof. A. Piazzoli
for constant interest in the work and support
and the International School for Advanced Studies, Trieste, 
where most of the work for this study has been done, 
for financial support.
The work of S.T.P. was supported in part by the EEC grant ERBFMRXCT960090
and by Grant PH-510 from the Bulgarian Science Foundation.



\newpage
\bec{\Large\bf{Figure Captions}}\eec

\noindent
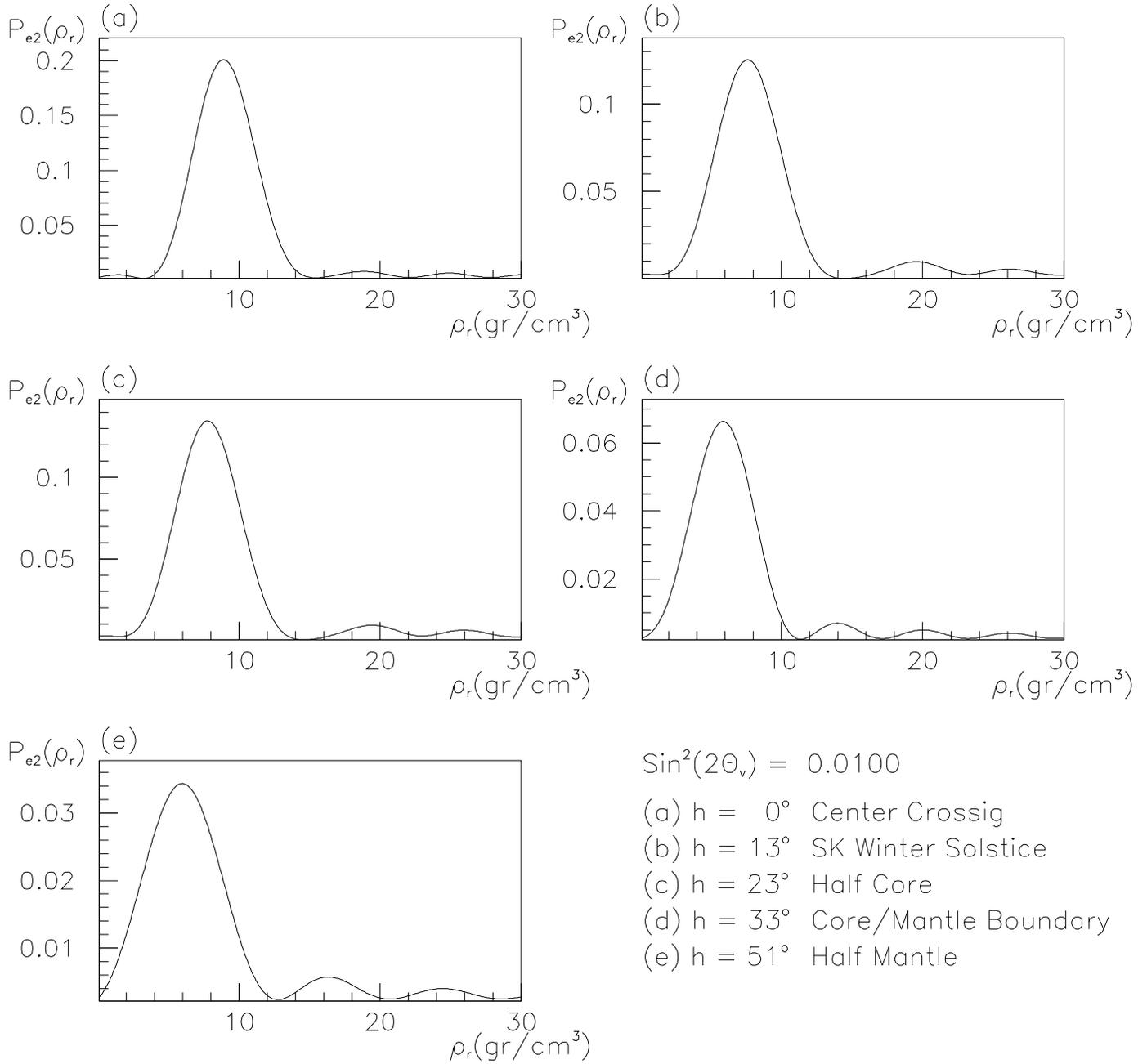
{\bf Figure 1.}
The dependence of the probability \PeTw\ on \rhoR\ for $\SdTvS = 0.01$. 
The five plots are obtained for 
$0.1 \mbox{ gr/cm}^3 \leq \rhoR \leq 30.0~\mbox{gr/cm}^3 $
($\Ye (mantle) = 0.49$, $\YeCore = 0.467$)
and five different solar neutrino trajectories in the Earth
determined by the Nadir angle \hath:
a) center crossing ($\hath = 0\degres$),
b) winter solstice for the \SK\ detector ($\hath = 13\deg$),
c) half core for the \SK\ detector ($\hath = 23\degres$),
d) core/mantle boundary ($\hath = 33\degres$),
e) half mantle ($\hath = 51\degres$).

\vspace{0.2cm}
\noindent
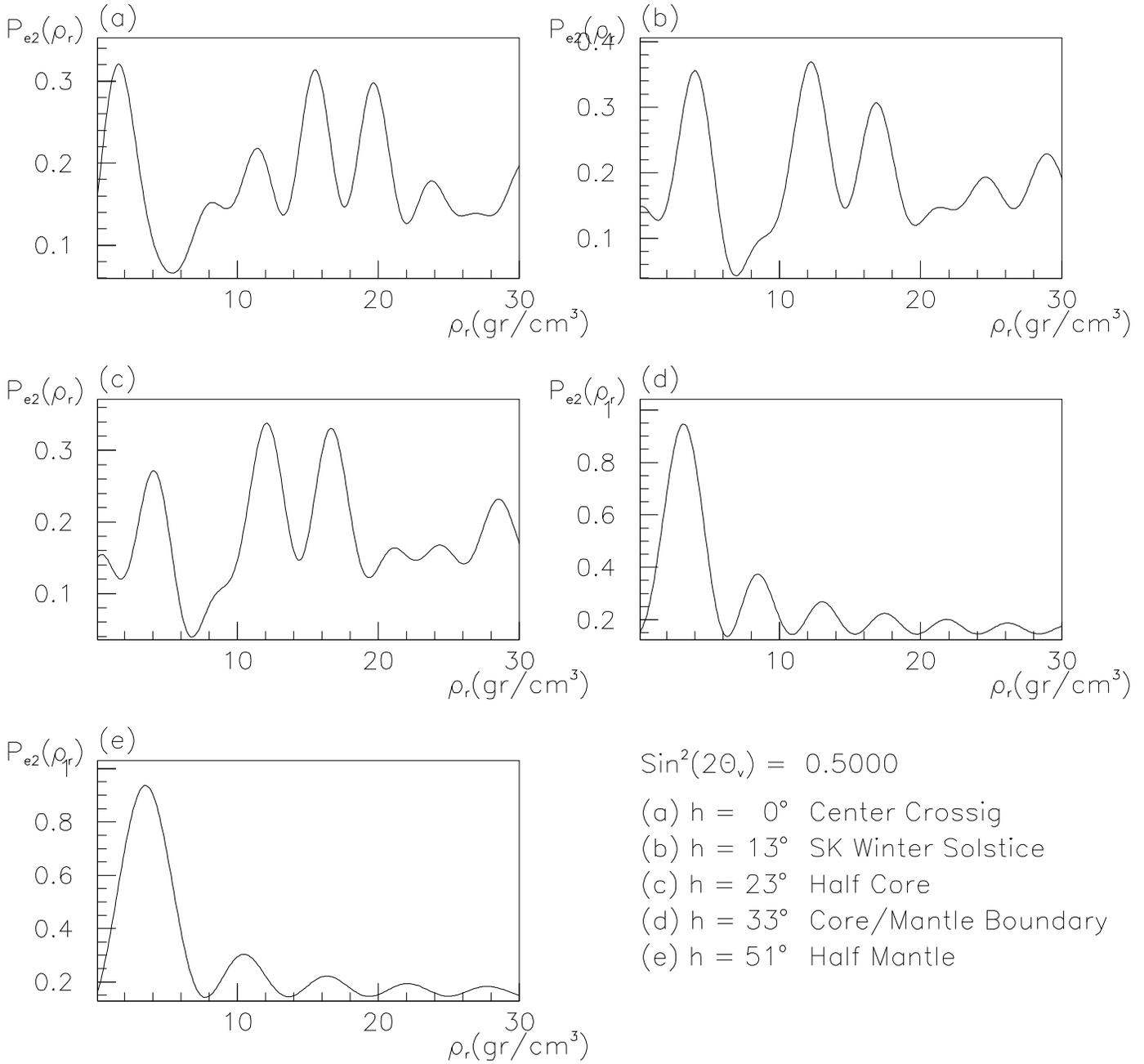
{\bf Figure 2.}
The dependence of the probability \PeTw\ on \rhoR\ for $\SdTvS = 0.5$.
The five plots were obtained for the same range of values of \rhoR\ and
neutrino trajectories as in figure 1.

\vspace{0.2cm}
\noindent
{\bf Figure 3.1 - 3.12} 
The probabilities \APeTw, \Ps and \PTot\ and the probability asymmetries
\AsymPs\ as functions of \rhoR\ and \EDms.
Each figure represents \APeTw, \Ps, \PTot\ and \AsymPs\ for the 
value of \SdTvS\ indicated in the figure.
The frames (a) show the probability \APeTw\ as a function of the 
density \rhoR\ for the \night\ ({\em short - dashed line}),
\mantle\ ({\em dotted line}) and \core\ ({\em solid line}) samples.
In all calculations \Ye\ in the mantle and in the core was assumed to be equal
respectively to 0.49 and 0.467.
The frames (b), (c) and (d) represent the probabilities (upper parts) 
\Ps\ ({\em dotted line}), 
\APeTw\ ({\em dashed line}) and \PTot\ ({\em solid line}), and the
probability asymmetry \AsymPs\ (lower parts)  
corresponding to the \night, \core\ and \mantle\ samples as functions of 
$\Enu/\dms$.
In some cases \APeTw\ is multiplied by a factor of $10$, which is indicated 
in the corresponding figure.

\vspace{0.2cm}
\noindent
{\bf{Figures 4.1 - 4.16.}}
Recoil - e$^{-}$ spectrum distortion $N^{s} (\Te)$ (upper frame) 
and D-N asymmetry in the spectrum \AsymSs\ (lower frame) 
(see eqs. 18 and 19) 
for the \DAY\ ({\em long-dashed line}), \night\ ({\em short-dashed line}),
\core\ ({\em solid line}) and \mantle\ ({\em dotted line}) samples
for $\Ye (mantle) = 0.49$, $\Ye (core) = 0.467$\ (Figs. 4.1 - 4.12) and
$\Ye (core) = 0.50$\ (Figs. 4.13 - 4.16).
The values of \dms\ and \SdTvS\ are indicated between the upper and the lower 
frames.

\vspace{0.2cm}
\noindent
{\bf{Figures 5a - 5b.}}
One year average event rate iso - (D-N) asymmetry contour plots for 
the \night\ sample for the \SK\ detector 
for $\TeTh = 5.0$\ (a) and 
$\TeTh = 7.5$\ MeV (b).
The {\em solid} ({\em dash-dotted}) lines correspond to 
$\Ye (core) = 0.467~(0.50)$.
The two light-grey areas limited by the dashed lines represent  
the ``conservative'' $\nu_e \rightarrow \nu_s$ solution regions, 
while the dark-grey area limited by the solid line
represents the reference solution region obtained 
in the second study quoted in ref. \cite{SPnu96}
within the standard solar model \cite{BP95}. 

\vspace{0.2cm}
\noindent
{\bf{Figures 6a - 6d.}}
One year average event rate iso - (D-N) asymmetry contour plots for 
the \core\ sample for the \SK\ detector:  figures (a) and (c) (b and d)
correspond to $\Ye (core) = 0.467$
($\Ye (core) = 0.5$), while figures (a) and (b) (c and d) were obtained
for $\TeTh = 5$\ MeV   
($\TeTh = 7.5$\ MeV).
The light-grey and dark-grey regions are the same 
as in Figs. 5a - 5b.

\vspace{0.2cm}
\noindent
{\bf{Figures 7a - 7b.}}
One year averaged event-rates iso - (D-N) asymmetry contour plots for 
the \mantle\ sample for the \SK\ detector:
figure (a) was obtained for $\TeTh = 5$\ MeV, 
while figure (b) corresponds to $\TeTh = 7.5$\ MeV.
The light-grey and dark-grey regions are the same as in Figs. 5a - 5b
and 6a - 6d.

\def\TDMSX{\mbox{$\times 10^{-6}$}}
\def\TDMFV{\mbox{$\times 10^{-5}$}}

 \newpage
 \def\TabEventRates{I}
 \begin{table}[ht]
 \bec
 \begin{tabular}{|rccc||cccc|}
 \multicolumn{8}{c}{\bf{Table \TabEventRates. 
 Event Rates for the \SK\ Detector}}\\
 \multicolumn{8}{c}{\bf{for $\YeCore = 0.467$, $\TeTh = 5$\ MeV.}}\\
 \hline
 \multicolumn{1}{|c}{N.}   & \SdTvS  & \dms\
     ${\rm [eV^2]}$   & $f_B$
     & \DAY     & \night 
     & \core   & \mantle \\
 \hline\hline
  1 & 0.001      & 3\TDMSX\   & 0.35 & 0.9188 & 0.9167 & 0.9107 & 0.9177 \\
  2 & 0.001      & 5\TDMSX\   & 0.35 & 0.8721 & 0.8718 & 0.8712 & 0.8719 \\
  3 & 0.001      & 7\TDMSX\   & 0.35 & 0.8274 & 0.8273 & 0.8273 & 0.8273 \\
 \hline                              
  4 & 0.002      & 3\TDMSX\   & 0.40 & 0.8446 & 0.8412 & 0.8311 & 0.8428 \\
  5 & 0.002      & 5\TDMSX\   & 0.40 & 0.7612 & 0.7608 & 0.7598 & 0.7610 \\
  6 & 0.002      & 7\TDMSX\   & 0.50 & 0.6858 & 0.6857 & 0.6856 & 0.6857 \\
  7 & 0.002      & 8\TDMSX\   & 0.50 & 0.6506 & 0.6505 & 0.6504 & 0.6505 \\
 \hline                              
  8 & 0.003      & 3\TDMSX\   & 0.50 & 0.7765 & 0.7723 & 0.7601 & 0.7743 \\
  9 & 0.003      & 5\TDMSX\   & 0.50 & 0.6646 & 0.6642 & 0.6630 & 0.6644 \\
 10 & 0.003      & 7\TDMSX\   & 0.50 & 0.5691 & 0.5690 & 0.5690 & 0.5690 \\
 \hline                              
 11 & 0.004      & 4\TDMSX\   & 0.50 & 0.6439 & 0.6422 & 0.6369 & 0.6431 \\
 12 & 0.004      & 5\TDMSX\   & 0.50 & 0.5809 & 0.5805 & 0.5795 & 0.5807 \\
 \hline                              
 13 & 0.005      & 4\TDMSX\   & 0.50 & 0.5775 & 0.5762 & 0.5716 & 0.5769 \\
 14 & 0.005      & 6\TDMSX\   & 0.70 & 0.4472 & 0.4473 & 0.4473 & 0.4473 \\
 15 & 0.005      & 7\TDMSX\   & 0.70 & 0.3937 & 0.3939 & 0.3940 & 0.3938 \\
 \hline                              
 16 & 0.007      & 3.3\TDMSX\ & 0.70 & 0.5270 & 0.5254 & 0.5206 & 0.5262 \\
 17 & 0.007      & 4\TDMSX\   & 0.70 & 0.4652 & 0.4649 & 0.4638 & 0.4651 \\
 18 & 0.007      & 5\TDMSX\   & 0.70 & 0.3896 & 0.3899 & 0.3901 & 0.3898 \\
 19 & 0.007      & 6\TDMSX\   & 1.00 & 0.3266 & 0.3270 & 0.3272 & 0.3270 \\
 \hline                              
 20 & 0.008      & 3\TDMSX\   & 0.70 & 0.5121 & 0.5108 & 0.5075 & 0.5113 \\
 21 & 0.008      & 4\TDMSX\   & 0.70 & 0.4178 & 0.4183 & 0.4195 & 0.4181 \\
 22 & 0.008      & 5\TDMSX\   & 1.00 & 0.3416 & 0.3421 & 0.3430 & 0.3420 \\
 23 & 0.008      & 6\TDMSX\   & 1.00 & 0.2796 & 0.2802 & 0.2805 & 0.2801 \\
 24 & 0.008      & 7\TDMSX\   & 1.50 & 0.2292 & 0.2297 & 0.2302 & 0.2296 \\
 \hline                              
 25 & 0.009      & 3\TDMSX\   & 0.70 & 0.4717 & 0.4718 & 0.4731 & 0.4716 \\
 26 & 0.009      & 4\TDMSX\   & 1.00 & 0.3755 & 0.3769 & 0.3808 & 0.3762 \\
 27 & 0.009      & 5\TDMSX\   & 1.00 & 0.2997 & 0.3007 & 0.3021 & 0.3004 \\
 28 & 0.009      & 6\TDMSX\   & 1.50 & 0.2397 & 0.2404 & 0.2408 & 0.2403 \\
 \hline                              
 29 & 0.010      & 3\TDMSX\   & 1.00 & 0.4345 & 0.4364 & 0.4429 & 0.4354 \\
 30 & 0.010      & 4\TDMSX\   & 1.00 & 0.3377 & 0.3400 & 0.3468 & 0.3389 \\
 31 & 0.010      & 5\TDMSX\   & 1.00 & 0.2633 & 0.2645 & 0.2668 & 0.2642 \\
 \hline                              
 32 & 0.014      & 4\TDMSX\   & 2.00 & 0.2221 & 0.2289 & 0.2493 & 0.2255 \\
 33 & 0.014      & 5\TDMSX\   & 2.00 & 0.1582 & 0.1611 & 0.1665 & 0.1601 \\
 \hline                              
 34 & 0.400      & 6\TDMSX\   & 3.00 & 0.1130 & 0.1649 & 0.1704 & 0.1640 \\
 35 & 0.400      & 8\TDMSX\   & 3.00 & 0.1133 & 0.1513 & 0.1798 & 0.1466 \\
 36 & 0.400      & 1\TDMFV\   & 3.00 & 0.1137 & 0.1393 & 0.1467 & 0.1381 \\
 \hline                              
 37 & 0.500      & 7\TDMSX\   & 2.00 & 0.1470 & 0.2017 & 0.2368 & 0.1960 \\
 38 & 0.500      & 8\TDMSX\   & 2.00 & 0.1471 & 0.1926 & 0.2222 & 0.1877 \\
 39 & 0.500      & 1\TDMFV\   & 2.00 & 0.1476 & 0.1787 & 0.1864 & 0.1774 \\
 \hline
 \end{tabular}
 \eec
 \end{table}

 \newpage
 \def\TabAsymmetries{II}
 \begin{table}[ht]
 \bec
 {\small
 \begin{tabular}{|rccc||rrrc||rrrc|}
 \multicolumn{12}{c}{\bf{Table \TabAsymmetries. 
 D - N Asymmetries for the \SK\ Detector}}\\
 \multicolumn{12}{c}{\bf{for $\TeTh = 5$\ MeV.}}\\
 \hline
 \hline
   \multicolumn{4}{|l||}{}&
   \multicolumn{4}{|c||}{$\YeCore = 0.467$}&
   \multicolumn{4}{|c|}{$\YeCore = 0.500$} \\
    \multicolumn{4}{|l||}{}
   &
   \multicolumn{3}{|c}{$\AsymRs \times 100$}
   &
   \multicolumn{1}{c||}
        {$\frac{|\AsymRC|}{|\AsymRN|}$}
   &
   \multicolumn{3}
        {|c}{$\AsymRs \times 100$}
   &
   \multicolumn{1}{c|}
          {$\frac{|\AsymRC|}{|\AsymRN|}$}
   \\
 \multicolumn{1}{|c}{N.}   &
        \SdTvS  & \dms   & $f_B$ &
 \multicolumn{1}{|c}{\night}   &
 \multicolumn{1}{c}{\core}      &
 \multicolumn{1}{c}{\mantle}    &
 \multicolumn{1}{c||}{}   &
 \multicolumn{1}{|c}{\night}      &
 \multicolumn{1}{c}{\core}      &
 \multicolumn{1}{c}{\mantle}    &
 \multicolumn{1}{c|}{}    \\
 \hline\hline
  1 & 0.001      & 3\TDMSX\   & 0.35 &  -0.22 &  -0.88 &  -0.12 &   3.97 &  -0.25 &  -1.08 &  -0.12 &   4.31 \\
  2 & 0.001      & 5\TDMSX\   & 0.35 &  -0.03 &  -0.11 &  -0.02 &   3.33 &  -0.07 &  -0.36 &  -0.02 &   5.25 \\
  3 & 0.001      & 7\TDMSX\   & 0.35 &  -0.01 &  -0.02 &  -0.01 &   1.60 &  -0.02 &  -0.06 &  -0.01 &   3.72 \\
 \hline                                                                                             
  4 & 0.002      & 3\TDMSX\   & 0.40 &  -0.41 &  -1.60 &  -0.21 &   3.95 &  -0.45 &  -1.93 &  -0.21 &   4.28 \\
  5 & 0.002      & 5\TDMSX\   & 0.40 &  -0.06 &  -0.19 &  -0.03 &   3.47 &  -0.11 &  -0.61 &  -0.03 &   5.37 \\
  6 & 0.002      & 7\TDMSX\   & 0.50 &  -0.02 &  -0.02 &  -0.01 &   1.52 &  -0.02 &  -0.09 &  -0.01 &   3.72 \\
  7 & 0.002      & 8\TDMSX\   & 0.50 &  -0.02 &  -0.03 &  -0.01 &   1.79 &  -0.02 &  -0.07 &  -0.01 &   3.71 \\
 \hline                                                                                             
  8 & 0.003      & 3\TDMSX\   & 0.50 &  -0.54 &  -2.13 &  -0.28 &   3.93 &  -0.60 &  -2.52 &  -0.28 &   4.23 \\
  9 & 0.003      & 5\TDMSX\   & 0.50 &  -0.07 &  -0.24 &  -0.04 &   3.63 &  -0.14 &  -0.74 &  -0.04 &   5.40 \\
 10 & 0.003      & 7\TDMSX\   & 0.50 &  -0.01 &  -0.01 &  -0.01 &   1.23 &  -0.02 &  -0.07 &  -0.01 &   3.71 \\
 \hline                                                                                             
 11 & 0.004      & 4\TDMSX\   & 0.50 &  -0.25 &  -1.09 &  -0.12 &   4.29 &  -0.37 &  -1.90 &  -0.12 &   5.16 \\
 12 & 0.004      & 5\TDMSX\   & 0.50 &  -0.06 &  -0.24 &  -0.03 &   3.77 &  -0.13 &  -0.71 &  -0.03 &   5.51 \\
 \hline                                                                                             
 13 & 0.005      & 4\TDMSX\   & 0.50 &  -0.24 &  -1.02 &  -0.11 &   4.35 &  -0.33 &  -1.68 &  -0.11 &   5.12 \\
 14 & 0.005      & 6\TDMSX\   & 0.70 &   0.01 &   0.02 &   0.01 &   1.80 &   0.01 &   0.02 &   0.01 &   1.32 \\
 15 & 0.005      & 7\TDMSX\   & 0.70 &   0.03 &   0.07 &   0.02 &   2.20 &   0.04 &   0.15 &   0.02 &   3.75 \\
 \hline                                                                                             
 16 & 0.007      & 3.3\TDMSX\ & 0.70 &  -0.31 &  -1.24 &  -0.16 &   3.94 &  -0.30 &  -1.18 &  -0.16 &   3.86 \\
 17 & 0.007      & 4\TDMSX\   & 0.70 &  -0.06 &  -0.30 &  -0.02 &   5.35 &  -0.02 &  -0.06 &  -0.02 &   2.77 \\
 18 & 0.007      & 5\TDMSX\   & 0.70 &   0.07 &   0.12 &   0.06 &   1.75 &   0.15 &   0.69 &   0.06 &   4.60 \\
 19 & 0.007      & 6\TDMSX\   & 1.00 &   0.11 &   0.18 &   0.10 &   1.69 &   0.19 &   0.73 &   0.10 &   3.95 \\
 \hline                                                                                             
 20 & 0.008      & 3\TDMSX\   & 0.70 &  -0.26 &  -0.90 &  -0.16 &   3.43 &  -0.20 &  -0.46 &  -0.16 &   2.28 \\
 21 & 0.008      & 4\TDMSX\   & 0.70 &   0.12 &   0.41 &   0.07 &   3.41 &   0.26 &   1.42 &   0.07 &   5.39 \\
 22 & 0.008      & 5\TDMSX\   & 1.00 &   0.17 &   0.41 &   0.13 &   2.43 &   0.36 &   1.74 &   0.13 &   4.84 \\
 23 & 0.008      & 6\TDMSX\   & 1.00 &   0.18 &   0.29 &   0.17 &   1.61 &   0.33 &   1.34 &   0.17 &   4.02 \\
 24 & 0.008      & 7\TDMSX\   & 1.50 &   0.23 &   0.43 &   0.20 &   1.87 &   0.35 &   1.25 &   0.20 &   3.58 \\
 \hline                                                                                             
 25 & 0.009      & 3\TDMSX\   & 0.70 &   0.04 &   0.31 &  -0.01 &   8.20 &   0.16 &   1.17 &  -0.01 &   7.33 \\
 26 & 0.009      & 4\TDMSX\   & 1.00 &   0.36 &   1.39 &   0.19 &   3.82 &   0.66 &   3.42 &   0.19 &   5.20 \\
 27 & 0.009      & 5\TDMSX\   & 1.00 &   0.31 &   0.81 &   0.23 &   2.57 &   0.64 &   3.14 &   0.23 &   4.89 \\
 28 & 0.009      & 6\TDMSX\   & 1.50 &   0.29 &   0.46 &   0.26 &   1.58 &   0.53 &   2.16 &   0.26 &   4.06 \\
 \hline                                                                                             
 29 & 0.010      & 3\TDMSX\   & 1.00 &   0.44 &   1.91 &   0.20 &   4.31 &   0.64 &   3.27 &   0.20 &   5.11 \\
 30 & 0.010      & 4\TDMSX\   & 1.00 &   0.69 &   2.68 &   0.35 &   3.90 &   1.17 &   5.98 &   0.35 &   5.10 \\
 31 & 0.010      & 5\TDMSX\   & 1.00 &   0.48 &   1.32 &   0.34 &   2.73 &   1.01 &   4.95 &   0.34 &   4.89 \\
 \hline                                                                                             
 32 & 0.014      & 4\TDMSX\   & 2.00 &   3.00 &  11.52 &   1.52 &   3.84 &   4.75 &  22.27 &   1.52 &   4.69 \\
 33 & 0.014      & 5\TDMSX\   & 2.00 &   1.78 &   5.10 &   1.21 &   2.86 &   3.66 &  17.18 &   1.21 &   4.70 \\
 \hline                                                                                             
 34 & 0.400      & 6\TDMSX\   & 3.00 &  37.34 &  40.50 &  36.81 &   1.08 &  40.05 &  57.23 &  36.81 &   1.43 \\
 35 & 0.400      & 8\TDMSX\   & 3.00 &  28.71 &  45.38 &  25.60 &   1.58 &  31.53 &  60.29 &  25.60 &   1.91 \\
 36 & 0.400      & 1\TDMFV\   & 3.00 &  20.23 &  25.36 &  19.37 &   1.25 &  21.13 &  31.19 &  19.37 &   1.48 \\
 \hline                                                                                             
 37 & 0.500      & 7\TDMSX\   & 2.00 &  31.41 &  46.79 &  28.57 &   1.49 &  33.87 &  60.07 &  28.57 &   1.77 \\
 38 & 0.500      & 8\TDMSX\   & 2.00 &  26.74 &  40.63 &  24.21 &   1.52 &  28.65 &  51.38 &  24.21 &   1.79 \\
 39 & 0.500      & 1\TDMFV\   & 2.00 &  19.09 &  23.26 &  18.38 &   1.22 &  19.76 &  27.72 &  18.38 &   1.40 \\
 \hline
 \end{tabular}
 } 
 \eec
 \end{table}

 \newpage
 \def\TabAsymmetriesMeV{III}
 \begin{table}[ht]
 \bec
 {\small
 \begin{tabular}{|rccc||rrrc||rrrc|}
 \multicolumn{12}{c}{\bf{Table \TabAsymmetriesMeV. 
 Threshold Energy Dependence
 of the D - N Asymmetries }}\\
 \multicolumn{12}{c}{\bf{for $\YeCore = 0.467$.}}\\
 \hline
 \hline
   \multicolumn{4}{|l||}{}&
   \multicolumn{4}{|c||}{$\TeTh = 5$ MeV}&
   \multicolumn{4}{|c|}{$\TeTh = 7.5$ MeV} \\
   \multicolumn{4}{|l||}{}
   &
   \multicolumn{3}{|c}{$\AsymRs \times 100$}
   &
   \multicolumn{1}{c||}
        {$\frac{|\AsymRC|}{|\AsymRN|}$}
   &
   \multicolumn{3}
        {|c}{$\AsymRs \times 100$}
   &
   \multicolumn{1}{c|}
          {$\frac{|\AsymRC|}{|\AsymRN|}$}
   \\
 \multicolumn{1}{|c}{N.}   &
        \SdTvS  & \dms   & $f_B$ &
 \multicolumn{1}{|c}{\night}   &
 \multicolumn{1}{c}{\core}      &
 \multicolumn{1}{c}{\mantle}    &
 \multicolumn{1}{c||}{}   &
 \multicolumn{1}{|c}{\night}      &
 \multicolumn{1}{c}{\core}      &
 \multicolumn{1}{c}{\mantle}    &
 \multicolumn{1}{c|}{}    \\
 \hline\hline
  1 & 0.001      & 3\TDMSX\   & 0.35 &  -0.22 &  -0.88 &  -0.12 &   3.97 &  -0.28 &  -1.06 &  -0.15 &   3.80 \\
  2 & 0.001      & 5\TDMSX\   & 0.35 &  -0.03 &  -0.11 &  -0.02 &   3.33 &  -0.05 &  -0.19 &  -0.03 &   3.95 \\
  3 & 0.001      & 7\TDMSX\   & 0.35 &  -0.01 &  -0.02 &  -0.01 &   1.60 &  -0.01 &  -0.02 &  -0.01 &   1.39 \\
 \hline                              
  4 & 0.002      & 3\TDMSX\   & 0.40 &  -0.41 &  -1.60 &  -0.21 &   3.95 &  -0.51 &  -1.92 &  -0.28 &   3.80 \\
  5 & 0.002      & 5\TDMSX\   & 0.40 &  -0.06 &  -0.19 &  -0.03 &   3.47 &  -0.08 &  -0.33 &  -0.04 &   3.88 \\
  6 & 0.002      & 7\TDMSX\   & 0.50 &  -0.02 &  -0.02 &  -0.01 &   1.52 &  -0.02 &  -0.02 &  -0.02 &   1.17 \\
  7 & 0.002      & 8\TDMSX\   & 0.50 &  -0.02 &  -0.03 &  -0.01 &   1.79 &  -0.02 &  -0.03 &  -0.01 &   1.97 \\
 \hline                              
  8 & 0.003      & 3\TDMSX\   & 0.50 &  -0.54 &  -2.13 &  -0.28 &   3.93 &  -0.68 &  -2.58 &  -0.37 &   3.79 \\
  9 & 0.003      & 5\TDMSX\   & 0.50 &  -0.07 &  -0.24 &  -0.04 &   3.63 &  -0.10 &  -0.41 &  -0.05 &   3.96 \\
 10 & 0.003      & 7\TDMSX\   & 0.50 &  -0.01 &  -0.01 &  -0.01 &   1.23 &  -0.01 &  -0.01 &  -0.01 &   1.02 \\
 \hline                              
 11 & 0.004      & 4\TDMSX\   & 0.50 &  -0.25 &  -1.09 &  -0.12 &   4.29 &  -0.39 &  -1.69 &  -0.18 &   4.34 \\
 12 & 0.004      & 5\TDMSX\   & 0.50 &  -0.06 &  -0.24 &  -0.03 &   3.77 &  -0.10 &  -0.42 &  -0.05 &   4.05 \\
 \hline                              
 13 & 0.005      & 4\TDMSX\   & 0.50 &  -0.24 &  -1.02 &  -0.11 &   4.35 &  -0.37 &  -1.59 &  -0.17 &   4.33 \\
 14 & 0.005      & 6\TDMSX\   & 0.70 &   0.01 &   0.02 &   0.01 &   1.80 &   0.00 &  -0.02 &   0.00 &   5.22 \\
 15 & 0.005      & 7\TDMSX\   & 0.70 &   0.03 &   0.07 &   0.02 &   2.20 &   0.02 &   0.05 &   0.02 &   2.44 \\
 \hline                              
 16 & 0.007      & 3.3\TDMSX\ & 0.70 &  -0.31 &  -1.24 &  -0.16 &   3.94 &  -0.50 &  -1.95 &  -0.27 &   3.86 \\
 17 & 0.007      & 4\TDMSX\   & 0.70 &  -0.06 &  -0.30 &  -0.02 &   5.35 &  -0.14 &  -0.58 &  -0.06 &   4.25 \\
 18 & 0.007      & 5\TDMSX\   & 0.70 &   0.07 &   0.12 &   0.06 &   1.75 &   0.05 &   0.10 &   0.04 &   2.03 \\
 19 & 0.007      & 6\TDMSX\   & 1.00 &   0.11 &   0.18 &   0.10 &   1.69 &   0.08 &   0.07 &   0.08 &   0.90 \\
 \hline                              
 20 & 0.008      & 3\TDMSX\   & 0.70 &  -0.26 &  -0.90 &  -0.16 &   3.43 &  -0.49 &  -1.72 &  -0.28 &   3.52 \\
 21 & 0.008      & 4\TDMSX\   & 0.70 &   0.12 &   0.41 &   0.07 &   3.41 &   0.09 &   0.41 &   0.04 &   4.44 \\
 22 & 0.008      & 5\TDMSX\   & 1.00 &   0.17 &   0.41 &   0.13 &   2.43 &   0.16 &   0.51 &   0.11 &   3.11 \\
 23 & 0.008      & 6\TDMSX\   & 1.00 &   0.18 &   0.29 &   0.17 &   1.61 &   0.14 &   0.14 &   0.14 &   1.00 \\
 24 & 0.008      & 7\TDMSX\   & 1.50 &   0.23 &   0.43 &   0.20 &   1.87 &   0.20 &   0.34 &   0.18 &   1.67 \\
 \hline                              
 25 & 0.009      & 3\TDMSX\   & 0.70 &   0.04 &   0.31 &  -0.01 &   8.20 &  -0.19 &  -0.54 &  -0.13 &   2.94 \\
 26 & 0.009      & 4\TDMSX\   & 1.00 &   0.36 &   1.39 &   0.19 &   3.82 &   0.41 &   1.75 &   0.18 &   4.30 \\
 27 & 0.009      & 5\TDMSX\   & 1.00 &   0.31 &   0.81 &   0.23 &   2.57 &   0.32 &   1.06 &   0.20 &   3.28 \\
 28 & 0.009      & 6\TDMSX\   & 1.50 &   0.29 &   0.46 &   0.26 &   1.58 &   0.23 &   0.24 &   0.23 &   1.06 \\
 \hline                              
 29 & 0.010      & 3\TDMSX\   & 1.00 &   0.44 &   1.91 &   0.20 &   4.31 &   0.23 &   1.02 &   0.09 &   4.52 \\
 30 & 0.010      & 4\TDMSX\   & 1.00 &   0.69 &   2.68 &   0.35 &   3.90 &   0.82 &   3.49 &   0.37 &   4.26 \\
 31 & 0.010      & 5\TDMSX\   & 1.00 &   0.48 &   1.32 &   0.34 &   2.73 &   0.53 &   1.79 &   0.32 &   3.38 \\
 \hline                              
 32 & 0.014      & 4\TDMSX\   & 2.00 &   3.00 &  11.52 &   1.52 &   3.84 &   3.68 &  14.82 &   1.71 &   4.03 \\
 33 & 0.014      & 5\TDMSX\   & 2.00 &   1.78 &   5.10 &   1.21 &   2.86 &   2.00 &   6.82 &   1.18 &   3.41 \\
 \hline                              
 34 & 0.400      & 6\TDMSX\   & 3.00 &  37.34 &  40.50 &  36.81 &   1.08 &  40.28 &  27.51 &  42.21 &   0.68 \\
 35 & 0.400      & 8\TDMSX\   & 3.00 &  28.71 &  45.38 &  25.60 &   1.58 &  34.15 &  57.48 &  29.53 &   1.68 \\
 36 & 0.400      & 1\TDMFV\   & 3.00 &  20.23 &  25.36 &  19.37 &   1.25 &  23.82 &  33.08 &  22.19 &   1.39 \\
 \hline                              
 37 & 0.500      & 7\TDMSX\   & 2.00 &  31.41 &  46.79 &  28.57 &   1.49 &  36.10 &  52.94 &  32.94 &   1.47 \\
 38 & 0.500      & 8\TDMSX\   & 2.00 &  26.74 &  40.63 &  24.21 &   1.52 &  31.72 &  51.71 &  27.88 &   1.63 \\
 39 & 0.500      & 1\TDMFV\   & 2.00 &  19.09 &  23.26 &  18.38 &   1.22 &  22.37 &  30.19 &  21.01 &   1.35 \\
 \hline
 \end{tabular}
 } 
 \eec
 \end{table}

 \newpage
 \def\TabAsymmetriesMeV{IV}
 \begin{table}[ht]
 \bec
 {\small
 \begin{tabular}{|rccc||rrrc||rrrc|}
 \multicolumn{12}{c}{\bf{Table \TabAsymmetriesMeV. 
 Threshold Energy Dependence
 of the D - N Asymmetries}}\\
  \multicolumn{12}{c}{\bf{for $\YeCore = 0.5$.}}\\
 \hline
 \hline
   \multicolumn{4}{|l||}{}&
   \multicolumn{4}{|c||}{$\TeTh = 5$ MeV}&
   \multicolumn{4}{|c|}{$\TeTh = 7.5$ MeV} \\
   \multicolumn{4}{|l||}{}
   &
   \multicolumn{3}{|c}{$\AsymRs \times 100$}
   &
   \multicolumn{1}{c||}
        {$\frac{|\AsymRC|}{|\AsymRN|}$}
   &
   \multicolumn{3}
        {|c}{$\AsymRs \times 100$}
   &
   \multicolumn{1}{c|}
          {$\frac{|\AsymRC|}{|\AsymRN|}$}
   \\
 \multicolumn{1}{|c}{N.}   &
        \SdTvS  & \dms   & $f_B$ &
 \multicolumn{1}{|c}{\night}   &
 \multicolumn{1}{c}{\core}      &
 \multicolumn{1}{c}{\mantle}    &
 \multicolumn{1}{c||}{}   &
 \multicolumn{1}{|c}{\night}      &
 \multicolumn{1}{c}{\core}      &
 \multicolumn{1}{c}{\mantle}    &
 \multicolumn{1}{c|}{}    \\
 \hline\hline
  1 & 0.001      & 3\TDMSX\   & 0.35 &  -0.25 &  -1.08 & -0.12 &   4.31 &  -0.28 &  -1.07 &  -0.15 &   3.82 \\
  2 & 0.001      & 5\TDMSX\   & 0.35 &  -0.07 &  -0.36 & -0.02 &   5.25 &  -0.10 &  -0.56 &  -0.03 &   5.50 \\
  3 & 0.001      & 7\TDMSX\   & 0.35 &  -0.02 &  -0.06 & -0.01 &   3.72 &  -0.02 &  -0.08 &  -0.01 &   3.80 \\
 \hline                                                                                            
  4 & 0.002      & 3\TDMSX\   & 0.40 &  -0.45 &  -1.93 & -0.21 &   4.28 &  -0.51 &  -1.94 &  -0.28 &   3.80 \\
  5 & 0.002      & 5\TDMSX\   & 0.40 &  -0.11 &  -0.61 & -0.03 &   5.37 &  -0.17 &  -0.96 &  -0.04 &   5.57 \\
  6 & 0.002      & 7\TDMSX\   & 0.50 &  -0.02 &  -0.09 & -0.01 &   3.72 &  -0.03 &  -0.11 &  -0.02 &   3.78 \\
  7 & 0.002      & 8\TDMSX\   & 0.50 &  -0.02 &  -0.07 & -0.01 &   3.71 &  -0.02 &  -0.09 &  -0.01 &   3.63 \\
 \hline                                                                                            
  8 & 0.003      & 3\TDMSX\   & 0.50 &  -0.60 &  -2.52 & -0.28 &   4.23 &  -0.68 &  -2.58 &  -0.37 &   3.79 \\
  9 & 0.003      & 5\TDMSX\   & 0.50 &  -0.14 &  -0.74 & -0.04 &   5.40 &  -0.21 &  -1.17 &  -0.05 &   5.60 \\
 10 & 0.003      & 7\TDMSX\   & 0.50 &  -0.02 &  -0.07 & -0.01 &   3.71 &  -0.03 &  -0.10 &  -0.01 &   3.83 \\
 \hline                                                                                            
 11 & 0.004      & 4\TDMSX\   & 0.50 &  -0.37 &  -1.90 & -0.12 &   5.16 &  -0.53 &  -2.68 &  -0.18 &   5.08 \\
 12 & 0.004      & 5\TDMSX\   & 0.50 &  -0.13 &  -0.71 & -0.03 &   5.51 &  -0.21 &  -1.15 &  -0.05 &   5.55 \\
 \hline                                                                                            
 13 & 0.005      & 4\TDMSX\   & 0.50 &  -0.33 &  -1.68 & -0.11 &   5.12 &  -0.49 &  -2.45 &  -0.17 &   5.04 \\
 14 & 0.005      & 6\TDMSX\   & 0.70 &   0.01 &   0.02 &  0.01 &   1.32 &  -0.02 &  -0.11 &   0.00 &   7.01 \\
 15 & 0.005      & 7\TDMSX\   & 0.70 &   0.04 &   0.15 &  0.02 &   3.75 &   0.03 &   0.12 &   0.02 &   4.02 \\
 \hline                                                                                            
 16 & 0.007      & 3.3\TDMSX\ & 0.70 &  -0.30 &  -1.18 & -0.16 &   3.86 &  -0.51 &  -2.02 &  -0.27 &   3.92 \\
 17 & 0.007      & 4\TDMSX\   & 0.70 &  -0.02 &  -0.06 & -0.02 &   2.77 &  -0.13 &  -0.57 &  -0.06 &   4.22 \\
 18 & 0.007      & 5\TDMSX\   & 0.70 &   0.15 &   0.69 &  0.06 &   4.60 &   0.13 &   0.68 &   0.04 &   5.16 \\
 19 & 0.007      & 6\TDMSX\   & 1.00 &   0.19 &   0.73 &  0.10 &   3.95 &   0.16 &   0.64 &   0.08 &   4.07 \\
 \hline                                                                                            
 20 & 0.008      & 3\TDMSX\   & 0.70 &  -0.20 &  -0.46 & -0.16 &   2.28 &  -0.45 &  -1.49 &  -0.28 &   3.27 \\
 21 & 0.008      & 4\TDMSX\   & 0.70 &   0.26 &   1.42 &  0.07 &   5.39 &   0.20 &   1.17 &   0.04 &   5.86 \\
 22 & 0.008      & 5\TDMSX\   & 1.00 &   0.36 &   1.74 &  0.13 &   4.84 &   0.38 &   2.02 &   0.11 &   5.32 \\
 23 & 0.008      & 6\TDMSX\   & 1.00 &   0.33 &   1.34 &  0.17 &   4.02 &   0.31 &   1.28 &   0.14 &   4.21 \\
 24 & 0.008      & 7\TDMSX\   & 1.50 &   0.35 &   1.25 &  0.20 &   3.58 &   0.33 &   1.22 &   0.18 &   3.69 \\
 \hline                                                                                            
 25 & 0.009      & 3\TDMSX\   & 0.70 &   0.16 &   1.17 & -0.01 &   7.33 &  -0.14 &  -0.24 &  -0.13 &   1.69 \\
 26 & 0.009      & 4\TDMSX\   & 1.00 &   0.66 &   3.42 &  0.19 &   5.20 &   0.66 &   3.50 &   0.18 &   5.30 \\
 27 & 0.009      & 5\TDMSX\   & 1.00 &   0.64 &   3.14 &  0.23 &   4.89 &   0.72 &   3.81 &   0.20 &   5.29 \\
 28 & 0.009      & 6\TDMSX\   & 1.50 &   0.53 &   2.16 &  0.26 &   4.06 &   0.50 &   2.14 &   0.23 &   4.26 \\
 \hline                                                                                            
 29 & 0.010      & 3\TDMSX\   & 1.00 &   0.64 &   3.27 &  0.20 &   5.11 &   0.28 &   1.40 &   0.09 &   5.00 \\
 30 & 0.010      & 4\TDMSX\   & 1.00 &   1.17 &   5.98 &  0.35 &   5.10 &   1.25 &   6.43 &   0.37 &   5.13 \\
 31 & 0.010      & 5\TDMSX\   & 1.00 &   1.01 &   4.95 &  0.34 &   4.89 &   1.16 &   6.09 &   0.32 &   5.26 \\
 \hline                                                                                            
 32 & 0.014      & 4\TDMSX\   & 2.00 &   4.75 &  22.27 &  1.52 &   4.69 &   5.29 &  24.50 &   1.71 &   4.63 \\
 33 & 0.014      & 5\TDMSX\   & 2.00 &   3.66 &  17.18 &  1.21 &   4.70 &   4.21 &  20.77 &   1.18 &   4.93 \\
 \hline                                                                                            
 34 & 0.400      & 6\TDMSX\   & 3.00 &  40.05 &  57.23 & 36.81 &   1.43 &  42.11 &  41.51 &  42.21 &   0.99 \\
 35 & 0.400      & 8\TDMSX\   & 3.00 &  31.53 &  60.29 & 25.60 &   1.91 &  38.17 &  76.15 &  29.53 &   2.00 \\
 36 & 0.400      & 1\TDMFV\   & 3.00 &  21.13 &  31.19 & 19.37 &   1.48 &  25.29 &  41.97 &  22.19 &   1.66 \\
 \hline                                                                                            
 37 & 0.500      & 7\TDMSX\   & 2.00 &  33.87 &  60.07 & 28.57 &   1.77 &  39.06 &  68.28 &  32.94 &   1.75 \\
 38 & 0.500      & 8\TDMSX\   & 2.00 &  28.65 &  51.38 & 24.21 &   1.79 &  34.46 &  65.56 &  27.88 &   1.90 \\
 39 & 0.500      & 1\TDMFV\   & 2.00 &  19.76 &  27.72 & 18.38 &   1.40 &  23.51 &  37.28 &  21.01 &   1.59 \\
 \hline                                                                                    
 \end{tabular}                                                                             
 } 
 \eec
 \end{table}

 \newpage
 \def\TabEventRates{V}
 \begin{table}[ht]
 \bec
 \begin{tabular}{|rccc||cccc|}
 \multicolumn{8}{c}{\bf{Table \TabEventRates. 
 Event Rates for the \SK\ Detector}}\\
 \multicolumn{8}{c}{\bf{for $\YeCore=0.5$, $\TeTh = 5$\ MeV.}}\\
 \hline
 \multicolumn{1}{|c}{N.}   & \SdTvS  & \dms\
     ${\rm [eV^2]}$   & $f_B$
     & \DAY     & \night 
     & \core   & \mantle \\
 \hline\hline
  1 & 0.001      & 3\TDMSX\   & 0.35 & 0.9188 & 0.9165 & 0.9090 &  0.9177\\
  2 & 0.001      & 5\TDMSX\   & 0.35 & 0.8721 & 0.8715 & 0.8690 &  0.8719\\
  3 & 0.001      & 7\TDMSX\   & 0.35 & 0.8274 & 0.8273 & 0.8269 &  0.8273\\
 \hline                                                                  
  4 & 0.002      & 3\TDMSX\   & 0.40 & 0.8446 & 0.8408 & 0.8285 &  0.8428\\
  5 & 0.002      & 5\TDMSX\   & 0.40 & 0.7612 & 0.7604 & 0.7566 &  0.7610\\
  6 & 0.002      & 7\TDMSX\   & 0.50 & 0.6858 & 0.6856 & 0.6852 &  0.6857\\
  7 & 0.002      & 8\TDMSX\   & 0.50 & 0.6506 & 0.6505 & 0.6502 &  0.6505\\
 \hline                                                                  
  8 & 0.003      & 3\TDMSX\   & 0.50 & 0.7765 & 0.7719 & 0.7572 &  0.7743\\
  9 & 0.003      & 5\TDMSX\   & 0.50 & 0.6646 & 0.6637 & 0.6597 &  0.6644\\
 10 & 0.003      & 7\TDMSX\   & 0.50 & 0.5691 & 0.5690 & 0.5687 &  0.5690\\
 \hline                                                                  
 11 & 0.004      & 4\TDMSX\   & 0.50 & 0.6439 & 0.6415 & 0.6318 &  0.6431\\
 12 & 0.004      & 5\TDMSX\   & 0.50 & 0.5809 & 0.5802 & 0.5768 &  0.5807\\
 \hline                                                                  
 13 & 0.005      & 4\TDMSX\   & 0.50 & 0.5775 & 0.5756 & 0.5679 &  0.5769\\
 14 & 0.005      & 6\TDMSX\   & 0.70 & 0.4472 & 0.4473 & 0.4473 &  0.4473\\
 15 & 0.005      & 7\TDMSX\   & 0.70 & 0.3937 & 0.3939 & 0.3943 &  0.3938\\
 \hline                                                                  
 16 & 0.007      & 3.3\TDMSX\ & 0.70 & 0.5270 & 0.5254 & 0.5209 &  0.5262\\
 17 & 0.007      & 4\TDMSX\   & 0.70 & 0.4652 & 0.4651 & 0.4649 &  0.4651\\
 18 & 0.007      & 5\TDMSX\   & 0.70 & 0.3896 & 0.3902 & 0.3923 &  0.3898\\
 19 & 0.007      & 6\TDMSX\   & 1.00 & 0.3266 & 0.3272 & 0.3290 &  0.3270\\
 \hline                                                                  
 20 & 0.008      & 3\TDMSX\   & 0.70 & 0.5121 & 0.5111 & 0.5098 &  0.5113\\
 21 & 0.008      & 4\TDMSX\   & 0.70 & 0.4178 & 0.4189 & 0.4238 &  0.4181\\
 22 & 0.008      & 5\TDMSX\   & 1.00 & 0.3416 & 0.3428 & 0.3475 &  0.3420\\
 23 & 0.008      & 6\TDMSX\   & 1.00 & 0.2796 & 0.2806 & 0.2834 &  0.2801\\
 24 & 0.008      & 7\TDMSX\   & 1.50 & 0.2292 & 0.2300 & 0.2321 &  0.2296\\
 \hline                                                                  
 25 & 0.009      & 3\TDMSX\   & 0.70 & 0.4717 & 0.4724 & 0.4772 &  0.4716\\
 26 & 0.009      & 4\TDMSX\   & 1.00 & 0.3755 & 0.3780 & 0.3886 &  0.3762\\
 27 & 0.009      & 5\TDMSX\   & 1.00 & 0.2997 & 0.3016 & 0.3093 &  0.3004\\
 28 & 0.009      & 6\TDMSX\   & 1.50 & 0.2397 & 0.2410 & 0.2449 &  0.2403\\
 \hline                                                                  
 29 & 0.010      & 3\TDMSX\   & 1.00 & 0.4345 & 0.4373 & 0.4490 &  0.4354\\
 30 & 0.010      & 4\TDMSX\   & 1.00 & 0.3377 & 0.3416 & 0.3585 &  0.3389\\
 31 & 0.010      & 5\TDMSX\   & 1.00 & 0.2633 & 0.2659 & 0.2766 &  0.2642\\
 \hline                                                                  
 32 & 0.014      & 4\TDMSX\   & 2.00 & 0.2221 & 0.2329 & 0.2778 &  0.2255\\
 33 & 0.014      & 5\TDMSX\   & 2.00 & 0.1582 & 0.1641 & 0.1879 &  0.1601\\
 \hline                                                                  
 34 & 0.400      & 6\TDMSX\   & 3.00 & 0.1130 & 0.1696 & 0.2037 &  0.1640\\
 35 & 0.400      & 8\TDMSX\   & 3.00 & 0.1133 & 0.1558 & 0.2111 &  0.1466\\
 36 & 0.400      & 1\TDMFV\   & 3.00 & 0.1137 & 0.1406 & 0.1557 &  0.1381\\
 \hline                                                                  
 37 & 0.500      & 7\TDMSX\   & 2.00 & 0.1470 & 0.2069 & 0.2732 &  0.1960\\
 38 & 0.500      & 8\TDMSX\   & 2.00 & 0.1471 & 0.1964 & 0.2489 &  0.1877\\
 39 & 0.500      & 1\TDMFV\   & 2.00 & 0.1476 & 0.1799 & 0.1951 &  0.1774\\
 \hline
 \end{tabular}
 \eec
 \end{table}

 \newpage
 \def\TabEventRates{VI}
 \begin{table}[ht]
 \bec
 \begin{tabular}{|rccc||cccc|}
 \multicolumn{8}{c}{\bf{Table \TabEventRates. 
 Event Rates for the \SK\ Detector}}\\
 \multicolumn{8}{c}{\bf{for $\YeCore=0.467$, $\TeTh=7.5$\ MeV.}}\\
 \hline
 \multicolumn{1}{|c}{N.}   & \SdTvS  & \dms\
     ${\rm [eV^2]}$   & $f_B$
     & \DAY     & \night 
     & \core   & \mantle \\
 \hline\hline
  1 & 0.001      & 3\TDMSX\   & 0.35 & 0.9284 & 0.9258 & 0.9187 & 0.9270 \\
  2 & 0.001      & 5\TDMSX\   & 0.35 & 0.8873 & 0.8869 & 0.8856 & 0.8871 \\
  3 & 0.001      & 7\TDMSX\   & 0.35 & 0.8479 & 0.8478 & 0.8478 & 0.8479 \\
 \hline                              
  4 & 0.002      & 3\TDMSX\   & 0.40 & 0.8623 & 0.8579 & 0.8458 & 0.8599 \\
  5 & 0.002      & 5\TDMSX\   & 0.40 & 0.7875 & 0.7869 & 0.7850 & 0.7872 \\
  6 & 0.002      & 7\TDMSX\   & 0.50 & 0.7194 & 0.7193 & 0.7192 & 0.7193 \\
  7 & 0.002      & 8\TDMSX\   & 0.50 & 0.6872 & 0.6871 & 0.6870 & 0.6871 \\
 \hline                              
  8 & 0.003      & 3\TDMSX\   & 0.50 & 0.8008 & 0.7954 & 0.7805 & 0.7979 \\
  9 & 0.003      & 5\TDMSX\   & 0.50 & 0.6989 & 0.6982 & 0.6961 & 0.6985 \\
 10 & 0.003      & 7\TDMSX\   & 0.50 & 0.6104 & 0.6103 & 0.6103 & 0.6103 \\
 \hline                              
 11 & 0.004      & 4\TDMSX\   & 0.50 & 0.6792 & 0.6766 & 0.6679 & 0.6781 \\
 12 & 0.004      & 5\TDMSX\   & 0.50 & 0.6205 & 0.6199 & 0.6180 & 0.6202 \\
 \hline                              
 13 & 0.005      & 4\TDMSX\   & 0.50 & 0.6170 & 0.6147 & 0.6072 & 0.6159 \\
 14 & 0.005      & 6\TDMSX\   & 0.70 & 0.4926 & 0.4925 & 0.4925 & 0.4925 \\
 15 & 0.005      & 7\TDMSX\   & 0.70 & 0.4403 & 0.4404 & 0.4405 & 0.4403 \\
 \hline                              
 16 & 0.007      & 3.3\TDMSX\ & 0.70 & 0.5684 & 0.5656 & 0.5574 & 0.5669 \\
 17 & 0.007      & 4\TDMSX\   & 0.70 & 0.5091 & 0.5084 & 0.5062 & 0.5087 \\
 18 & 0.007      & 5\TDMSX\   & 0.70 & 0.4351 & 0.4353 & 0.4355 & 0.4353 \\
 19 & 0.007      & 6\TDMSX\   & 1.00 & 0.3721 & 0.3724 & 0.3723 & 0.3724 \\
 \hline                              
 20 & 0.008      & 3\TDMSX\   & 0.70 & 0.5539 & 0.5512 & 0.5445 & 0.5524 \\
 21 & 0.008      & 4\TDMSX\   & 0.70 & 0.4626 & 0.4630 & 0.4645 & 0.4628 \\
 22 & 0.008      & 5\TDMSX\   & 1.00 & 0.3868 & 0.3874 & 0.3887 & 0.3872 \\
 23 & 0.008      & 6\TDMSX\   & 1.00 & 0.3236 & 0.3241 & 0.3241 & 0.3241 \\
 24 & 0.008      & 7\TDMSX\   & 1.50 & 0.2709 & 0.2715 & 0.2719 & 0.2714 \\
 \hline                              
 25 & 0.009      & 3\TDMSX\   & 0.70 & 0.5148 & 0.5138 & 0.5120 & 0.5141 \\
 26 & 0.009      & 4\TDMSX\   & 1.00 & 0.4205 & 0.4222 & 0.4279 & 0.4213 \\
 27 & 0.009      & 5\TDMSX\   & 1.00 & 0.3439 & 0.3451 & 0.3476 & 0.3446 \\
 28 & 0.009      & 6\TDMSX\   & 1.50 & 0.2817 & 0.2823 & 0.2823 & 0.2823 \\
 \hline                              
 29 & 0.010      & 3\TDMSX\   & 1.00 & 0.4784 & 0.4795 & 0.4833 & 0.4789 \\
 30 & 0.010      & 4\TDMSX\   & 1.00 & 0.3823 & 0.3854 & 0.3959 & 0.3837 \\
 31 & 0.010      & 5\TDMSX\   & 1.00 & 0.3060 & 0.3076 & 0.3115 & 0.3070 \\
 \hline                              
 32 & 0.014      & 4\TDMSX\   & 2.00 & 0.2619 & 0.2717 & 0.3038 & 0.2664 \\
 33 & 0.014      & 5\TDMSX\   & 2.00 & 0.1926 & 0.1965 & 0.2062 & 0.1949 \\
 \hline                              
 34 & 0.400      & 6\TDMSX\   & 3.00 & 0.1129 & 0.1699 & 0.1490 & 0.1734 \\
 35 & 0.400      & 8\TDMSX\   & 3.00 & 0.1131 & 0.1597 & 0.2044 & 0.1524 \\
 36 & 0.400      & 1\TDMFV\   & 3.00 & 0.1134 & 0.1441 & 0.1584 & 0.1417 \\
 \hline                              
 37 & 0.500      & 7\TDMSX\   & 2.00 & 0.1468 & 0.2115 & 0.2525 & 0.2047 \\
 38 & 0.500      & 8\TDMSX\   & 2.00 & 0.1469 & 0.2023 & 0.2494 & 0.1946 \\
 39 & 0.500      & 1\TDMFV\   & 2.00 & 0.1472 & 0.1843 & 0.1996 & 0.1818 \\
 \hline
 \end{tabular}
 \eec
 \end{table}

 \newpage
 \def\TabEventRates{VII}
 \begin{table}[ht]
 \bec
 \begin{tabular}{|rccc||cccc|}
 \multicolumn{8}{c}{\bf{Table \TabEventRates. 
 Event Rates for the \SK\ Detector}}\\
 \multicolumn{8}{c}{\bf{
  for $\YeCore=0.5$, $\TeTh=7.5$\ MeV.}}\\
 \hline
 \multicolumn{1}{|c}{N.}   & \SdTvS  & \dms\
     ${\rm [eV^2]}$   & $f_B$
     & \DAY     & \night 
     & \core   & \mantle \\
 \hline\hline
  1 & 0.001      & 3\TDMSX\   & 0.35 & 0.9284 & 0.9258 & 0.9185 &  0.9270\\
  2 & 0.001      & 5\TDMSX\   & 0.35 & 0.8873 & 0.8864 & 0.8823 &  0.8871\\
  3 & 0.001      & 7\TDMSX\   & 0.35 & 0.8479 & 0.8478 & 0.8473 &  0.8479\\
 \hline                                                                  
  4 & 0.002      & 3\TDMSX\   & 0.40 & 0.8623 & 0.8579 & 0.8457 &  0.8599\\
  5 & 0.002      & 5\TDMSX\   & 0.40 & 0.7875 & 0.7862 & 0.7800 &  0.7872\\
  6 & 0.002      & 7\TDMSX\   & 0.50 & 0.7194 & 0.7192 & 0.7186 &  0.7193\\
  7 & 0.002      & 8\TDMSX\   & 0.50 & 0.6872 & 0.6870 & 0.6866 &  0.6871\\
 \hline                                                                  
  8 & 0.003      & 3\TDMSX\   & 0.50 & 0.8008 & 0.7954 & 0.7804 &  0.7979\\
  9 & 0.003      & 5\TDMSX\   & 0.50 & 0.6989 & 0.6974 & 0.6908 &  0.6985\\
 10 & 0.003      & 7\TDMSX\   & 0.50 & 0.6104 & 0.6102 & 0.6097 &  0.6103\\
 \hline                                                                  
 11 & 0.004      & 4\TDMSX\   & 0.50 & 0.6792 & 0.6757 & 0.6613 &  0.6781\\
 12 & 0.004      & 5\TDMSX\   & 0.50 & 0.6205 & 0.6193 & 0.6135 &  0.6202\\
 \hline                                                                  
 13 & 0.005      & 4\TDMSX\   & 0.50 & 0.6170 & 0.6140 & 0.6020 &  0.6159\\
 14 & 0.005      & 6\TDMSX\   & 0.70 & 0.4926 & 0.4925 & 0.4920 &  0.4925\\
 15 & 0.005      & 7\TDMSX\   & 0.70 & 0.4403 & 0.4404 & 0.4408 &  0.4403\\
 \hline                                                                  
 16 & 0.007      & 3.3\TDMSX\ & 0.70 & 0.5684 & 0.5655 & 0.5571 &  0.5669\\
 17 & 0.007      & 4\TDMSX\   & 0.70 & 0.5091 & 0.5084 & 0.5062 &  0.5087\\
 18 & 0.007      & 5\TDMSX\   & 0.70 & 0.4351 & 0.4357 & 0.4381 &  0.4353\\
 19 & 0.007      & 6\TDMSX\   & 1.00 & 0.3721 & 0.3727 & 0.3745 &  0.3724\\
 \hline                                                                  
 20 & 0.008      & 3\TDMSX\   & 0.70 & 0.5539 & 0.5514 & 0.5457 &  0.5524\\
 21 & 0.008      & 4\TDMSX\   & 0.70 & 0.4626 & 0.4635 & 0.4681 &  0.4628\\
 22 & 0.008      & 5\TDMSX\   & 1.00 & 0.3868 & 0.3882 & 0.3947 &  0.3872\\
 23 & 0.008      & 6\TDMSX\   & 1.00 & 0.3236 & 0.3246 & 0.3278 &  0.3241\\
 24 & 0.008      & 7\TDMSX\   & 1.50 & 0.2709 & 0.2718 & 0.2743 &  0.2714\\
 \hline                                                                  
 25 & 0.009      & 3\TDMSX\   & 0.70 & 0.5148 & 0.5140 & 0.5135 &  0.5141\\
 26 & 0.009      & 4\TDMSX\   & 1.00 & 0.4205 & 0.4233 & 0.4354 &  0.4213\\
 27 & 0.009      & 5\TDMSX\   & 1.00 & 0.3439 & 0.3464 & 0.3573 &  0.3446\\
 28 & 0.009      & 6\TDMSX\   & 1.50 & 0.2817 & 0.2831 & 0.2878 &  0.2823\\
 \hline                                                                  
 29 & 0.010      & 3\TDMSX\   & 1.00 & 0.4784 & 0.4798 & 0.4852 &  0.4789\\
 30 & 0.010      & 4\TDMSX\   & 1.00 & 0.3823 & 0.3871 & 0.4077 &  0.3837\\
 31 & 0.010      & 5\TDMSX\   & 1.00 & 0.3060 & 0.3096 & 0.3252 &  0.3070\\
 \hline                                                                  
 32 & 0.014      & 4\TDMSX\   & 2.00 & 0.2619 & 0.2762 & 0.3351 &  0.2664\\
 33 & 0.014      & 5\TDMSX\   & 2.00 & 0.1926 & 0.2009 & 0.2373 &  0.1949\\
 \hline                                                                  
 34 & 0.400      & 6\TDMSX\   & 3.00 & 0.1129 & 0.1732 & 0.1721 &  0.1734\\
 35 & 0.400      & 8\TDMSX\   & 3.00 & 0.1131 & 0.1665 & 0.2523 &  0.1524\\
 36 & 0.400      & 1\TDMFV\   & 3.00 & 0.1134 & 0.1463 & 0.1737 &  0.1417\\
 \hline                                                                  
 37 & 0.500      & 7\TDMSX\   & 2.00 & 0.1468 & 0.2181 & 0.2991 &  0.2047\\
 38 & 0.500      & 8\TDMSX\   & 2.00 & 0.1469 & 0.2081 & 0.2903 &  0.1946\\
 39 & 0.500      & 1\TDMFV\   & 2.00 & 0.1472 & 0.1865 & 0.2147 &  0.1818\\
 \hline
 \end{tabular}
 \eec
 \end{table}


\begin{thebibliography}{200}

\bibitem{ArticleI}{
M. Maris and S.T. Petcov, Phys. Rev. D {\bf 56}, 7444 (1997).
}

\bibitem{ArticleII}{
Q.Y. Liu, M. Maris and S.T. Petcov,
Phys. Rev. D {\bf 56}, 5991 (1997).
}

\bibitem{SPnu96} {
P.I. Krastev and S.T. Petcov, reported by 
S.T. Petcov in {\it ``Neutrino '96''}, 
Proceedings of the Int. Conference on Neutrino Physics and Astrophysics,
June 13 - 19, 1996, Helsinki, Finland 
(eds. K. Enqvist, K. Huitu and J. Maalampi,
World Scientific, Singapore, 1997), p. 106;
J.N. Bahcall, P.I. Krastev and A.Yu. Smirnov, 
work in progress;
S.T. Petcov, 
in Proceedings of the 4th Int. Conference on Solar Neutrinos,
April 8 - 11, 1997, Heidelberg, Germany 
(ed. W. Hampel, Max-Planck-Institut fur
Kernphysik, Heidelberg, 1997), p. 309.
}


\bibitem{KPUNPUB96} {P.I. Krastev and S.T. Petcov, unpublished.
}


\bibitem{KS94} {P.I. Krastev and A.Yu. Smirnov, 
Phys. Lett. 338B (1994) 282;
V. Berezinsky, G. Fiorentini and M. Lissia,
Phys. Lett. B {\bf 341}, 38 (1995); N. Hata and P. Langacker, Phys. Rev. 
D {\bf 52}, 420
(1995).
}


\bibitem{Baltz:Weneser:1994}{
A.J. Baltz and J. Weneser,
Phys. Rev. D {\bf 50}, 5971 (1994).
}

\bibitem{Hata:Langacker:1994Earth}{
N. Hata and P. Langacker,
Phys. Rev. D {\bf 50}, 632 (1994).
}

\bibitem{Gelb:Kwong:Rosen:1996}{
J.M. Gelb, W. Kwong and S.P. Rosen, Phys. Rev. Lett. {\bf 78}, 2296 (1997).

}

\bibitem{Lisi:Montanino:1997}{
E. Lisi and D. Montanino, Phys. Rev. D {\bf 56}, 1792 (1997).
}

\bibitem{Bahcall:Krastev:1997}{
J.N. Bahcall and P.I. Krastev, Phys. Rev. C {\bf 56}, 2839 (1997).
}

\bibitem{Hata:1997}{N. Hata, Talk given at the Conference on
Solar Neutrinos: News About SNUs,
December 2 - 6, 1997, Institute of Theoretical Physics,
University of California, Santa Barbara, U. S. A.
}

\bibitem{NU4DN} {M. Maris, in Proceedings of the 
4th Int. Conference on Solar Neutrinos,
April 8 - 11, 1997, Heidelberg, Germany 
(ed. W. Hampel, Max-Planck-Institut fur
Kernphysik, Heidelberg, 1997), p. 342.
 
}

\bibitem{SKDNII:spectrum}{M. Maris and S.T. Petcov, 
 E-archive report hep-ph/9703207.}


\bibitem{BP95}{
J.N. Bahcall and M.H. Pinsonneault,
Rev. Mod. Phys. {\bf 67}, 781 (1995).
}

\bibitem{KPNPB95}
{ P.I. Krastev and S.T. Petcov, Nucl. Phys. B {\bf 449}, 605 (1995).
}


\bibitem{KPL96}  {P.I. Krastev, Q.Y. Liu and S.T. Petcov, 
Phys. Rev. D {\bf 54}, 7057 (1996).
}


\bibitem{KamDN} {K.S. Hirata et al., 
Phys. Rev. D {\bf 44}, 2241 (1991) and 
 Phys. Rev. Lett. {\bf 66}, 9 (1991).

}

\bibitem{SKSB97} {R. Svoboda (Super-Kamiokande collaboration),
 Talk given at the Conference on
Solar Neutrinos: News About SNUs,
December 2 - 6, 1997, Institute of Theoretical Physics,
University of California, Santa Barbara, U. S. A.
}

\bibitem{Langacker:1986} {P. Langacker et al., Nucl. Phys. 
B {\bf 282}, 589 (1986).
}

\bibitem{Stacey:1977}{
F.D. Stacey,
 {\it Physics of the Earth, 2$^{nd}$ edition},
John Wiley and Sons, London, New York, 1977.

}

\bibitem{PREM81}{A.D. Dziewonski and D.L. Anderson, 
               Physics of the Earth and Planetary Interiors 
               {\bf 25}, 297 (1981).
}

\bibitem{CORE}{R. Jeanloz,
Annu. Rev. Earth Planet. Sci. {\bf{18}}, 356 (1990);
C.J. All{\`e}gre et al., Earth
Planet. Sci. Lett. {\bf 134}, 515 (1995);
W.F. McDonough and S.-s. Sun, Chemical Geology {\bf 120}, 223 (1995).
}

\bibitem{Bahcall:etal:1996}{
J.N. Bahcall et al.,
Phys. Rev. D {\bf{54}}, 411 (1996).

}

\bibitem{Bahcall:Kamionkowski:Sirlin:1995}{
J.N. Bahcall, M. Kamionkowski and A. Sirlin,
Phys. Rev. D {\bf{51}}, 6146 (1995).

}

\bibitem{QLAS97} {Q.Y. Liu and A.Yu. Smirnov, E-archive report 
hep-ph/9712493.
}

\bibitem{KWROSEN96} {W. Kwong and S.P. Rosen, 
Phys. Rev. D{\bf 51}, 6159 (1995).
}

\bibitem{BilGiunt95} {S.M. Bilenky and C. Giunti,
Phys. Lett. B{\bf 320}, 323 (1994); Z. Phys.
C{\bf 68}, 495 (1995).
}

\bibitem{Maris:Petcov:1998} {M. Maris and S.T. Petcov, in preparation.
}

\end{thebibliography}
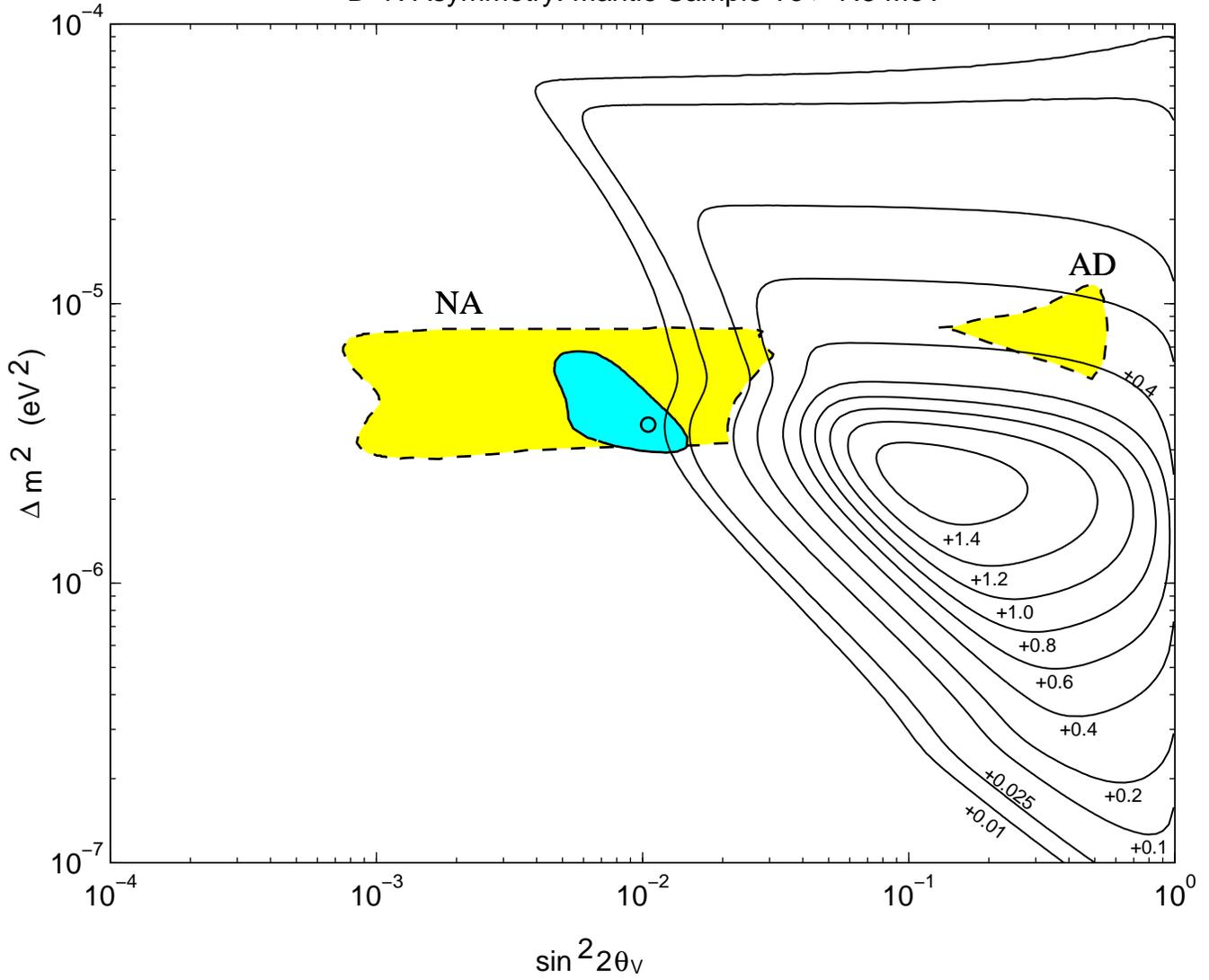
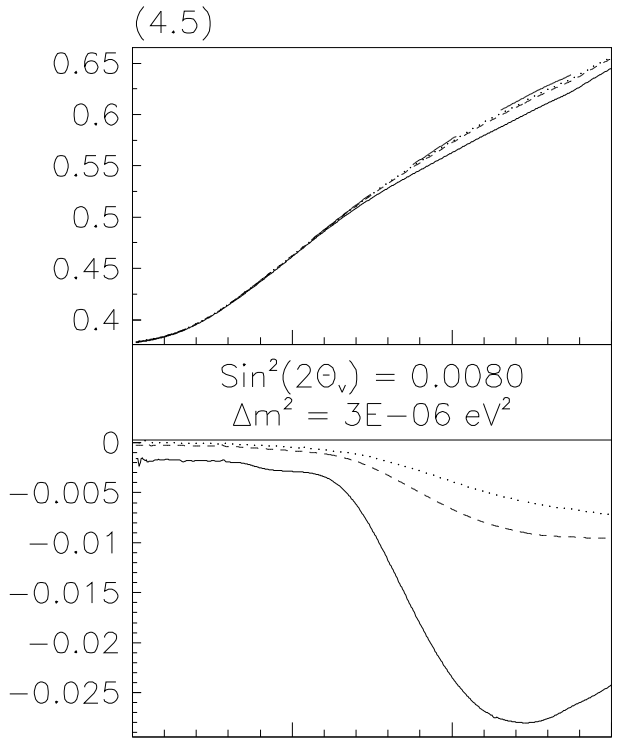
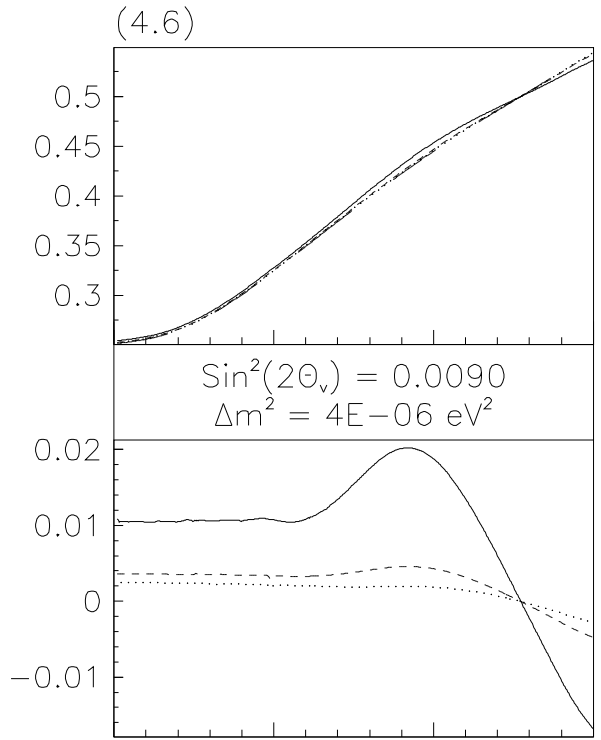
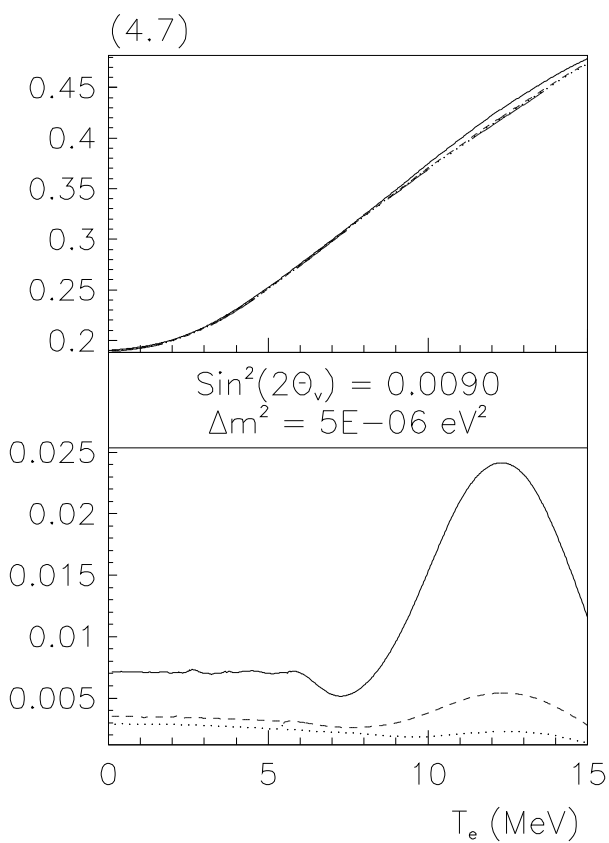
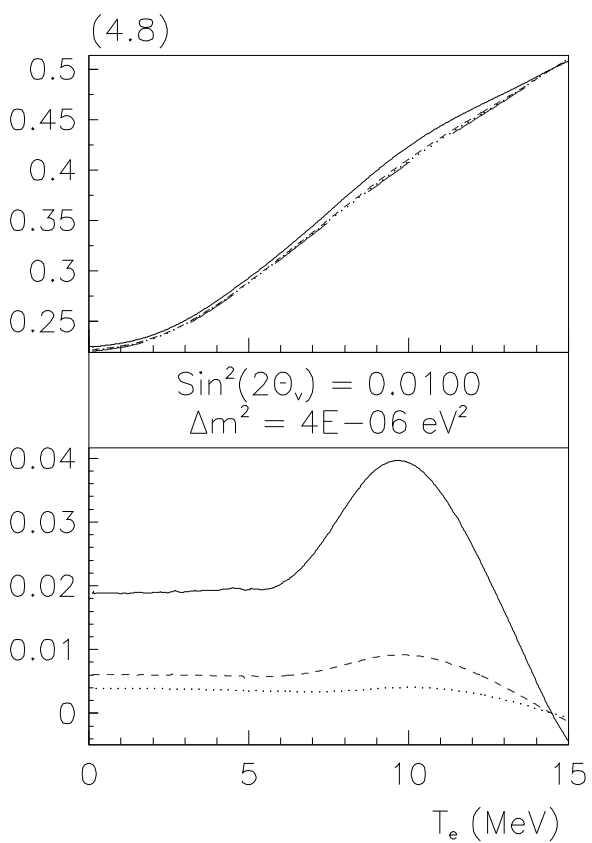
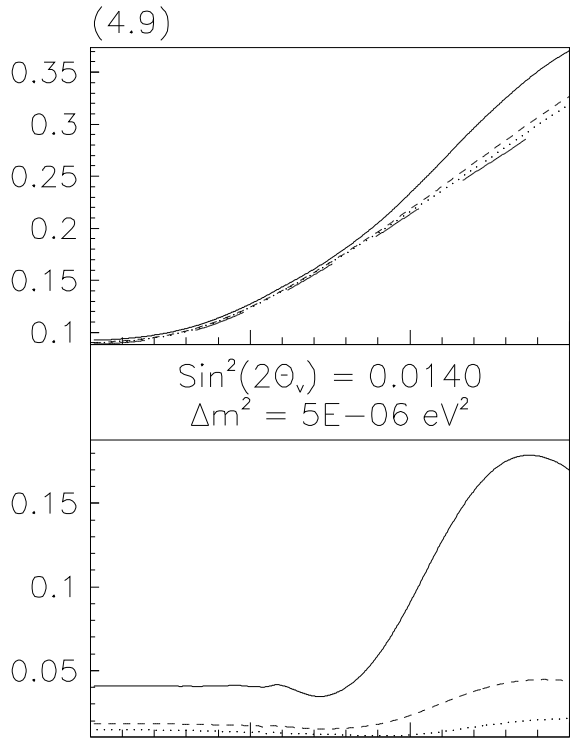
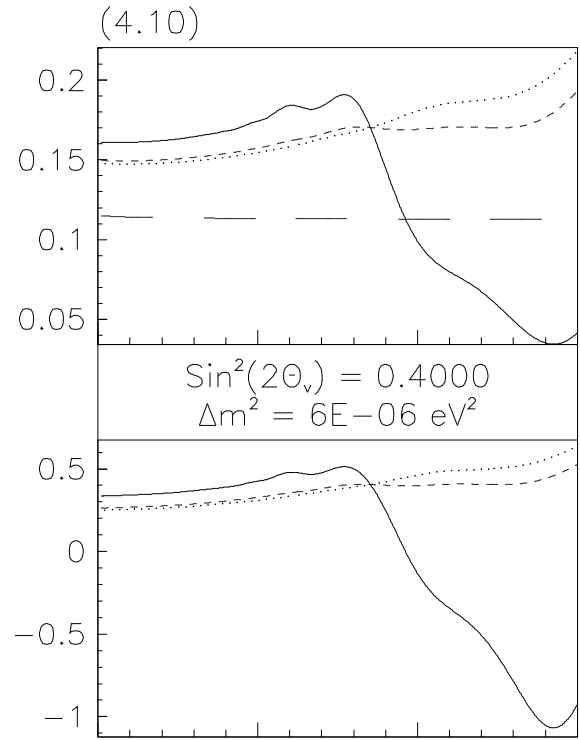
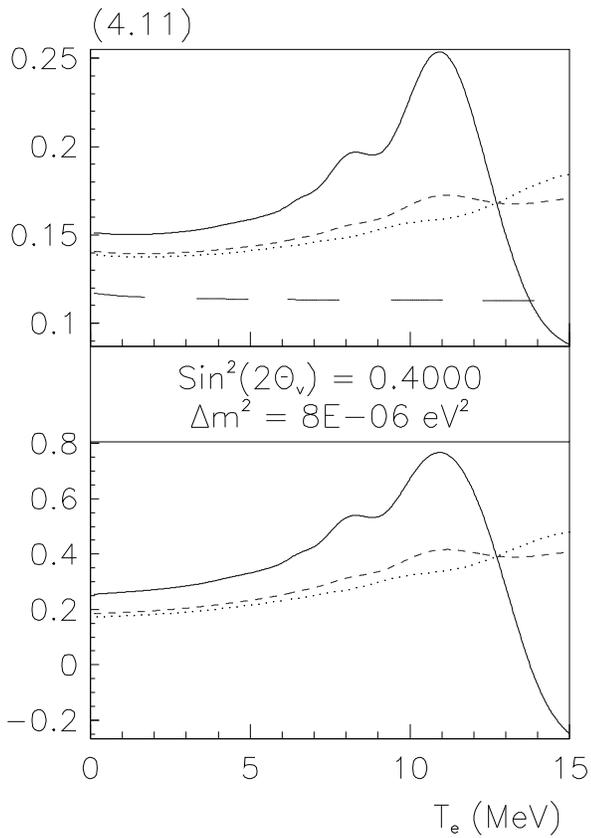
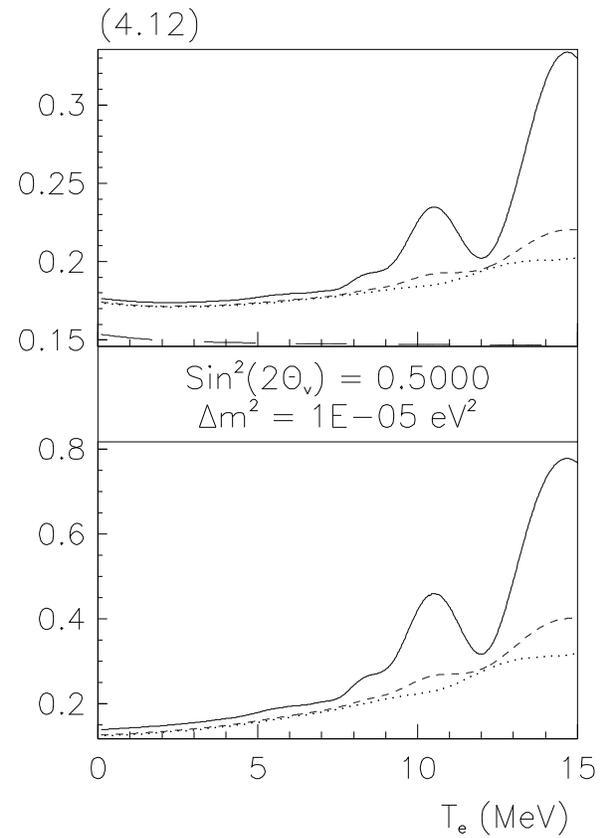
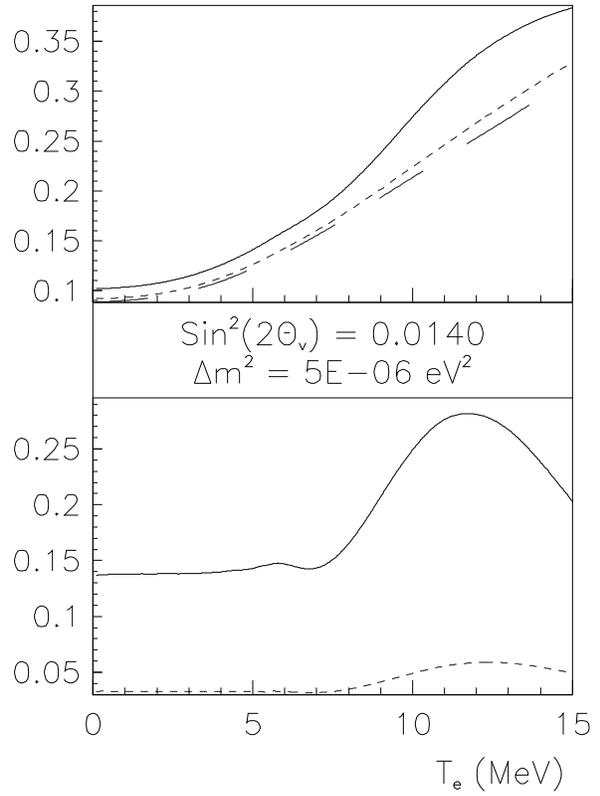
\end{document}